%% file: main.tex
\documentclass[resetfootnote,twocolumn]{aastex7}

\usepackage{tikz}
\usetikzlibrary{shapes.geometric, arrows}

\tikzstyle{startstop} = [rectangle, rounded corners, 
minimum width=3cm, 
minimum height=1cm,
text centered, 
draw=black, 
fill=red!30]

\tikzstyle{io} = [trapezium, 
trapezium stretches=true, 
trapezium left angle=70, 
trapezium right angle=110, 
minimum width=3cm, 
minimum height=1cm, text centered, 
text width=5cm, 
draw=black, fill=blue!30]

\tikzstyle{process} = [rectangle, 
minimum width=3cm, 
minimum height=1cm, 
text centered, 
text width=3cm, 
draw=black, 
fill=orange!30]

\tikzstyle{process2} = [rectangle, 
minimum width=3cm, 
minimum height=1cm, 
text centered, 
text width=5cm, 
draw=black, 
fill=orange!30]

\tikzstyle{decision} = [diamond, 
minimum width=3cm, 
minimum height=1cm, 
text centered, 
draw=black, 
fill=green!30]
\tikzstyle{arrow} = [thick,->,>=stealth]

\usepackage{easyReview}
\usepackage{fontawesome}
\usepackage{hyperref}
\usepackage{xcolor}
\usepackage{mhchem}
\usepackage{amsmath}
\usepackage{caption}
\usepackage{subcaption}
\usepackage{multirow}
\usepackage{comment}

\newcommand{\Rey}{\text{Re}}
\newcommand{\Kn}{\text{Kn}}
\newcommand{\Cs}{\beta}

\submitjournal{AAS Journal}

\shorttitle{\texttt{Virga}}
\shortauthors{Batalha et al.}

\begin{document}

\title{Condensation Clouds in Substellar Atmospheres  with \texttt{Virga}}

\correspondingauthor{Natasha E. Batalha}
\email{natasha.e.batalha@nasa.gov}

\author[0000-0003-1240-6844]{Natasha E. Batalha}
\affiliation{NASA Ames Research Center}
\email{natasha.e.batalha@nasa.gov}

\author{Caoimhe M. Rooney}
\affiliation{NASA Ames Research Center}
\email{}

\author[0000-0001-6627-6067]{Channon Visscher}
\affiliation{Chemistry \& Planetary Sciences, Dordt University, Sioux Center, IA}
\affiliation{Center for Extrasolar Planetary Systems, Space Science Institute, Boulder, CO}
\email{}

\author[0000-0002-6721-3284]{Sarah E. Moran}
\altaffiliation{NHFP Sagan Fellow}
\affiliation{NASA Goddard Space Flight Center}
\email{sarah.e.moran@nasa.gov}

\author{Mark S. Marley}
\affiliation{NASA Ames Research Center}
\email{}

\author[0000-0002-5669-035X]{Aditya R. Sengupta}
\affiliation{Department of Astronomy and Astrophysics, University of California, Santa Cruz}
\email{}


\author{Sven Kiefer}
\affiliation{Department of Astronomy, University of Texas at Austin, 2515 Speedway, Austin, TX 78712, USA}
\email{}

\author[0000-0002-9733-0617]{Matt G. Lodge}
\affiliation{School of Physics, University of Bristol, Bristol, UK}
\email{}

\author[0000-0001-5864-9599]{James Mang}
\altaffiliation{NSF Graduate Research Fellow.}
\affiliation{Department of Astronomy, University of Texas at Austin, 2515 Speedway, Austin, TX 78712, USA}
\email{}

\author{Caroline V. Morley}
\affiliation{Department of Astronomy, University of Texas at Austin, 2515 Speedway, Austin, TX 78712, USA}
\email{}

\author{Sagnick Mukherjee}
\affiliation{Department of Astronomy and Astrophysics, University of California, Santa Cruz}
\email{}


\author[0000-0002-9843-4354]{Jonathan J. Fortney}
\affiliation{Department of Astronomy and Astrophysics, University of California, Santa Cruz}
\email{}

\author[0000-0002-8518-9601]{Peter Gao}
\affiliation{Earth and Planets Laboratory, Carnegie Institution for Science, 5241 Broad Branch Road, NW, Washington, DC 20015, USA}
\email{}

\author[0000-0002-8507-1304]{Nikole K. Lewis}
\affiliation{Department of Astronomy and Carl Sagan Institute, Cornell University, 122 Sciences Drive, Ithaca, NY 14853, USA}
\email{}

\author[0000-0002-4321-4581]{L. C. Mayorga}
\affiliation{Johns Hopkins Applied Physics Laboratory, 11100 Johns Hopkins Rd, Laurel, MD, 20723, USA}
\email{laura.mayorga@jhuapl.edu}

\author[0000-0003-3904-7378]{Logan A. Pearce}
\affil{Department of Astronomy, University of Michigan, Ann Arbor, MI 48109, USA}
\email{lapearce@umich.edu}

\author[0000-0003-4328-3867]{Hannah R. Wakeford}
\affiliation{University of Bristol, HH Wills Physics Laboratory, Tyndall Avenue, Bristol, UK}
\email{}



\begin{abstract}
Here we present an open-source cloud model for substellar atmospheres, called \texttt{Virga}.  The \texttt{Virga-v0} series has already been widely adopted in the literature. It is written in Python and has heritage from the \citet{ackerman2001cloud} model (often referred to as \texttt{eddysed}), used to study clouds on both exoplanets and brown dwarfs. In the development of the official \texttt{Virga-v1} we have retained all the original functionality of \texttt{eddysed} and updated/expanded several components including the back-end optical constants data, calculations of the Mie properties, available condensate species, saturation vapor pressure curves and formalism for fall speeds calculations. Here we benchmark \texttt{Virga} by reproducing key results in the literature, including the SiO$_2$ cloud detection in WASP-17 b and the brown dwarf  Diamondback-Sonora model series. Development of \texttt{Virga} is ongoing, with future versions already planned and ready for release. We encourage community feedback and collaborations within the GitHub code repository. 
\end{abstract}



\section{Introduction} \label{sec:intro}

Though atmospheres are primarily characterized by their gas-phase compositions, aerosols suspended within atmospheres fundamentally shape -- and are shaped by -- the chemistry, dynamics, and structure of atmospheres across all scales. The observable spectra, photometry, and visible appearances of brown dwarfs, Solar system worlds, and exoplanets are all influenced by these aerosols. ``Aerosol'' is a generic term indicating any sort of non-gaseous solid or liquid particle suspended in the atmosphere, with the more specific terms ``cloud'' having come to refer to a condensate and ``haze'' referring to a photochemical solid \citep[e.g.,][]{Gao2021JGRE..12606655G}. Both sub-types of aerosol can exist in most atmospheres, but   condensate clouds are far better understood both from a microphysical formation and chemical standpoint, based on equilibrium thermodynamic theory \citep[e.g.,][]{Lodders1998book,Visscher2010atmospheric}. Here, we focus on condensate clouds -- which we'll simply refer to as ``clouds'' through the rest of this work -- and their formation in all manner of substellar worlds. 

There are many closed- and a few open-source models relevant to the study of clouds in substellar objects, with varying degrees of enforced physics. There are Mie codes which compute scattering and extinction coefficients given optical constants and particle radii (e.g., \texttt{PyMieScatt} \citep{sumlin2018}, \texttt{MiePython} \citep{miepython}, \texttt{Optool} \citep{optool}, \texttt{LX-MIE} \citep{Kitzmann2018optical}). There are radiative transfer models that parameterize clouds in various ways to compute spectra, given a condensate, particle radius, and set of Mie properties (either computed internally or using pre-computed Mie properties). For example, Brewster \citep{brewster}, petitRADTRANS \citep{Molliere2019,prtRetrievals}, PLATON \citep{platon}, POSEIDON \citep{poseidon,Mullens2024}, and PICASO \citep{picaso} all have functionality to implement this purely ``flexible cloud'' approach. This approach does not physically restrict what condensates are allowed to form in the atmosphere, nor at what pressure-level. Therefore, it is especially useful for fitting spectroscopic data of substellar objects. However, in many contexts it is also useful to understand what species are expected to condense, at what pressure--temperature level, and with what particle radius.

As described in more detail in Section 1 of \citet{ackerman2001cloud}, the original need for a substellar cloud model was motivated by the unmistakable presence of cloud layers in brown dwarfs. Efforts to model the emerging L- to T-type spectral sequence with dusty atmospheres lacking discrete cloud layers of finite thickness led to arbitrarily red (in near-IR spectral slope and colors) models that could not reproduce the spectral changes seen across the L- to T-type transition. Adding finite thickness clouds allowed the models to have the correct qualitative behavior \citep[][]{Marley2000} but without an underlying process-based model. Clearly, a physically-motivated 1D cloud model was needed to understand the substellar clouds. This was the void that was addressed by the development of the original \texttt{Eddysed} model by \citet[][referred to as AM01 throughout the text]{ackerman2001cloud}. 

The aim of Ackerman and Marley was to create a simplified cloud model governed by a few free parameters that could predict the vertical distribution of condensates, without being computationally prohibitive. Prior to his work on AM01, Ackerman along with Eric Jensen at NASA Ames had generalized the CARMA cloud microphysics code \citep{carma1,carma2} originally developed by Brian Toon and Richard Turco \citep{Toon1988} to better model sedimentation and coagulation of particles. Ackerman brought his experience in this development to the project of creating a simpler 1D cloud model for astrophysical applications. AM01 was a foundational paper whose methodology has been used in hundreds of manuscripts for brown dwarfs \citep[e.g.,][]{marley2002clouds,Stephens2009ApJ...702..154S,morley2012neglected}, transiting planets \citep[e.g.,][]{Fortney2005MNRAS.364..649F,Morley2015ApJ...815..110M,Gao2020NatAs...4..951G}, and directly imaged planets \citep[e.g.,][]{Marley2012ApJ...754..135M}.
Though the original Fortran code was never officially released, AM01 was open-sourced as a Python code with the beta release of \texttt{Virga-v0} in 2019.  

As an aside, though we do not cover the subject of microphysical models in depth, we note that the methodology of AM01 today provides a conceptual steppingstone to these more complex codes such as CARMA \cite[e.g.,][]{carma1, carma2,Gao2018ApJ...855...86G, Powell2018ApJ...860...18P} and DRIFT (or StaticWeather)
\cite[e.g.,][]{Helling2001A&A...376..194H,Witte2009A&A...506.1367W,Witte2011A&A...529A..44W, Helling2004A&A...423..657H,Helling2006A&A...455..325H, Helling2013RSPTA.37110581H,Helling2017A&A...603A.123H,Woitke2020A&A...634A..23W}. For a full review of the full range of cloud methodologies see \citet{Gao2021JGRE..12606655G}.

Despite the lack of an official publication, \texttt{Virga-v0} \citep[Zenodo v0][]{2020zndo...3759888B} has already been widely used in the literature to study exoplanets and brown dwarfs. It has been used to study clouds in exoplanets using JWST \citep[e.g.,][]{Grant+2023, Inglis2024ApJ...973L..41I,caleb2025arXiv250206966C},  HST \citep[e.g.,][]{Boehm2025AJ....169...23B}, and Kepler photometry \citep{Hamill2024ApJ...976..181H}. It has been used to study clouds more generally in substellar objects \citep[e.g.,][]{Madurowicz2023AJ....165..238M,Chen2024MNRAS.533.3114C,Moran2024ApJ...973L...3M}. It has also been used at a population-level to study samples of planets \cite[e.g., six hot Jupiters' optical albedos][]{2022ApJ...926..157A}. Furthermore, it has been benchmarked in the literature with the microphysical code \texttt{CARMA} \citep{Mang2024ApJ...974..190M}. 

The core functionality of \texttt{Virga} is described in AM01. Here, we aim to clarify some of the original manuscript, while connecting it to the workflow of the \texttt{Virga} package. We also describe updates made to the original workflow. Ultimately, herein constitutes the first  official version released as \texttt{v1} on Zenodo \citep{virga1}.

In what follows we first describe the general code workflow and methodology in \S \ref{sec:virga}. Then we dive into additional modeling specifics in \S \ref{sec:details}, including a discussion of the optical constants data, the Mie grid, condensate formation model, eddy diffusion coefficient, and fall speed calculations. Finally, we end with a benchmark calculation for exoplanets and brown dwarfs in \S \ref{sec:benchE} and \S \ref{sec:benchB}, respectively, before concluding in \S \ref{sec:discon}. 

\section{\texttt{Virga}: The Code Workflow } \label{sec:virga}

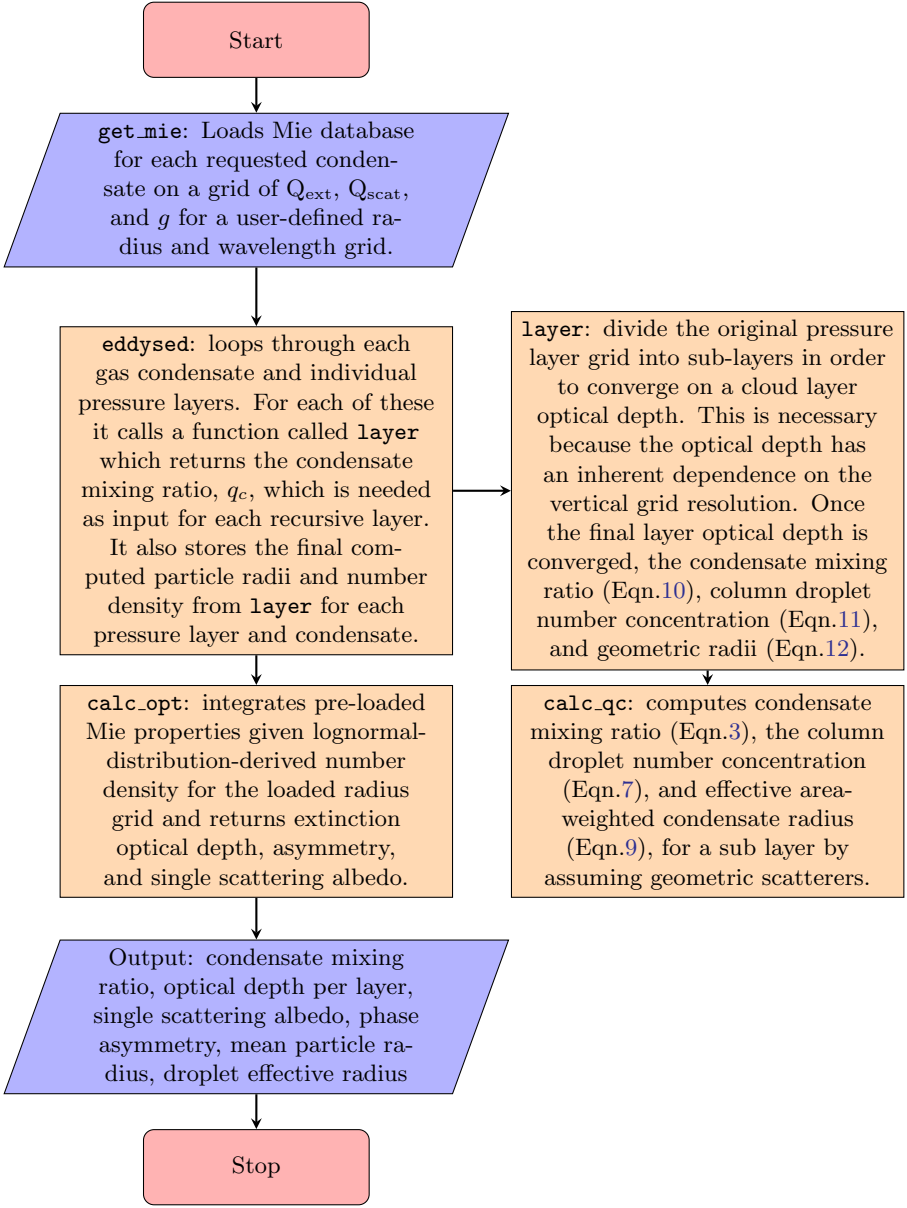
\begin{figure*}[htb]
\input{diagram.tex}
\caption{Top level \texttt{Virga} workflow showing the major functions of the code and their respective functionality.}
\label{fig:diagram}
\end{figure*}

In this section, we specifically focus on connecting the top-level methodology to the workflow of the  \texttt{Virga} package. The foundational equation that is most frequently highlighted when discussing AM01 is that which describes the balance between the upward turbulent mixing of condensate and vapor material (left, below) by the downward transport of condensate caused by sedimentation (right, below): 
\begin{equation}
    -K_\mathrm{zz} \frac{\delta q_t}{\delta z} - f_\mathrm{sed} w_* q_c = 0.
\label{eqn:balance}
\end{equation}
Here, $K_\mathrm{zz}$ is the eddy diffusion parameter (cm$^2$/sec), $q_t$ is the total condensate ($q_c$) plus vapor ($q_v$) mixing ratio, $z$ is the altitude, and $w_*$ is the convective velocity scale ($=K_\mathrm{zz}/L$ where $L$ is the mixing length). The term $f_{\mathrm sed}$ was introduced as a sedimentation efficiency parameter\footnote{AM01 initially used $f_{\rm rain}$ for this term but it was revised to remove any implication of liquid water.} controlling the vertical thickness of the clouds and defined as the ratio of the mass-weighted droplet sedimentation velocity to $w_*$.
While not appearing directly in this equation, the condensate particle radii (which along with $q_c$ are a desired output product) are influenced by the choice of $f_{\rm sed}$.   

To solve Eqn. \ref{eqn:balance} along with the particle radii,  \texttt{Virga} uses a top-level function called \texttt{eddysed}, which triggers two more embedded functions shown in the workflow of Figure \ref{fig:diagram}. \texttt{eddysed} loops through gas condensates and individual pressure layers while calling the function, \texttt{layer}. \texttt{layer} computes a few pertinent atmospheric parameters such as mean free path, number density, and atmospheric viscosity. Then, it sub-divides the user-defined atmosphere pressure grid into sub-grid layers and calls \texttt{calc\_qc} for each of these sub-layers. This sub-division is necessary because the optical depth has an inherent dependence on the vertical grid resolution. 

In  \texttt{calc\_qc}, in pursuit of solving Eqn. \ref{eqn:balance}, we first start by computing the saturation vapor pressure, $P_{v,i}$, at a given sub-layer, $i$, that has a corresponding temperature ($T_i$) and pressure ($P_i$). We describe the methodology to derive $P_{v,i}$ fully in \S \ref{sec:cloudcond}. From $P_{v,i}$ we compute the mass mixing ratio of the saturated vapor (g/g): 
\begin{equation}
    q_{v,i} =  \frac{f_s P_{v,i}}{R_c T_i \rho_a}
\end{equation}
where $f_s=S+1$ and $S$ is a user-defined parameter corresponding to the potential supersaturation prior to condensation (default, $S=0$). $R_c$ is the specific gas condensate for the given cloud species ($=R/\mu_c$ in units erg/K/g), and $\rho_a$ is the atmospheric density given the ideal gas law. $q_{v,i}$ is then compared to the total vapor+condensate mixing ratio in the immediately preceding layer towards higher pressure, $q_\mathrm{below}$. We describe the chemical calculations that we use for the very first condensing cloud base layer concentrations in \S \ref{sec:cloudcond}. If $q_\mathrm{below}>q_{v,i}$  we proceed with solving Eqn. \ref{eqn:balance}. Otherwise we assume $q_c=0$ such that any rain evaporation below the cloud is not accounted for. Ultimately, the solution takes the form 
\begin{equation}
    q_{t,i} = q_{v,i} + (q_\mathrm{below} - q_{v,i})  \exp(-f_\mathrm{sed} * \Delta z / L)
\label{eqn:qt}
\end{equation}
\begin{equation}
    q_{c,i} = \max(0,q_{t,i}-q_{v,i})
\end{equation}
if we require all excess vapor to condense and assume constant-with-altitude $q_c/q_t$. Here, the mixing length is $L$ and is described in detail in \S \ref{sec:kzz}. 

Next, we proceed within \texttt{calc\_qc} to compute the fall particle radius, $r_w$. The fall particle radius is computed by finding the radius which balances the fall speed of a spherical particle and the convective velocity scale: 
\begin{equation}
    v_f (r_w) - w_* = 0 
\label{eqn:rbalance}
\end{equation}
We detail our method of calculating fall speed in \S \ref{sec:fallspeed}. To arrive at a solution for the fall radius, we use \texttt{scipy.optimize.root\_scalar} with the \texttt{brentq} method, which implements the classic Brent's method to find a root of the function in a sign changing interval. We emphasize that  $w_*=K_\mathrm{zz}/L$, making the fall particle radius highly dependent on $K_\mathrm{zz}$. Within \texttt{Virga} there are a few different ways of either computing or user-defining $K_\mathrm{zz}$, which we detail in \S \ref{sec:kzz}. From the fall speed radius, we ultimately want to obtain the geometric mean radius of the lognormal size distribution. The fall radius, $r_w$, is related to the geometric mean radius, $r_g$ via the equation derived in AM01: 
\begin{equation}
    r_{g,i} = r_{w,i} f_\mathrm{sed}^{1/\alpha_i} \exp ( -\frac{\alpha_i + 6 }{2} \ln^2 \sigma_g )
\label{eqn:rgi}
\end{equation}
with the associated column droplet number concentration (cm$^{-2}$):
\begin{equation}
    N_{dz_{i}} =
    \frac{3 \rho_{a,i} q_{c,i}}{4 \pi \rho_p r_{g,i}^3} \exp(-\frac{9}{2} \ln^2\sigma_g) dz_i .
    \label{eqn:ndzi}
\end{equation} 
This introduces two new parameters: $\alpha$, the power-law dependence for the particle fall-speed about the solution to Eqn. \ref{eqn:rbalance}, and $\sigma_g$, the user-defined log normal particle distribution width. The power-law approximation to obtain $\alpha$ is described in detail in \S \ref{sec:fallspeed}. 

Note Eqn. \ref{eqn:ndzi}, computed within \texttt{calc\_qc} is still computed at the sub-layer, $i$ atmospheric level. We do this ensure the computed layer optical depth is insensitive to the resolution of the pressure grid, with layer thickness defined as $\Delta p$. The outer \texttt{Virga} function, \texttt{layer}, recursively splits the pressure grid into $n_\mathrm{sub}$ sub-layers until the   layer optical depth, computed via 
\begin{equation}
    \Delta \tau = \sum_{i=1}^{n_\mathrm{sub}} \frac{3}{2}\frac{q_{c,i}\Delta p_i}{g \rho_p r_{\textrm{eff},i}},
    \label{eqn:deltatau}
\end{equation}
is converged to the 1\%-level.  Here $r_{\textrm{eff},i}$ is area-weighted particle radius, referred to as the effective radius, $g$ is the gravity. The effective radius is related to geometric particle radius, $r_{g,i}$ (Eqn. \ref{eqn:rgi}) via
\begin{equation}
    r_{\textrm{eff},i} = r_{g,i} \exp \left( {\frac{5}{2} \ln^2\sigma_g} \right).
    \label{eqn:reffi}
\end{equation}
Note this effective (area-weighted) particle radius can also be expressed via the fall-speed radius, $r_w$, and sedimentation efficiency, $f_\textrm{sed}$, as in AM01 Eqn. 17.

After \texttt{layer} converges on a $\Delta \tau$ (Eqn. \ref{eqn:deltatau}) we can  compute the final layer averaged properties of interest: condensate mixing ratio ($q_c$), column droplet number concentration ($N_{dz}$), and geometric particle radius ($r_g$).
\begin{equation}
    q_{c/t} = \sum_{i=1}^{n_\mathrm{sub}} q_{c/t,i} \frac{\Delta p_i}{g} 
    \label{eqn:qc}
\end{equation}
\begin{equation}
    N_{dz} = \sum_{i=1}^{n_\mathrm{sub}} N_{dz_{i}}
    \label{eqn:ndz}
\end{equation}
\begin{equation}
    r_\textrm{g} = \frac{3}{2} \frac{q_c}{\rho_p \Delta \tau} \exp \left({-\frac{5}{2} \ln^2\sigma_g} \right).
    \label{eqn:rg}
\end{equation}
Note the latter relationship for $r_g$, Eqn. \ref{eqn:rg}, is derived by first computing the layer average area-weighted effective particle radius, $r_\textrm{eff}$ via the relationship in Eqn. \ref{eqn:deltatau} and then inverting Eqn.  \ref{eqn:reffi}. 

Tracking the flow backward of Figure \ref{fig:diagram} we note that function \texttt{eddysed} loops through this process for each condensate species, $s$ and altitude level, $z$. Inherently this means that all our computed physical properties are independent of other condensate species in the system (e.g. $q_c$ from Eqn.  \ref{eqn:qc} becomes $q_c(s,z)$) 

Now that all the condensate and particle properties are computed, we move on to get the final optical and scattering properties needed in radiative transfer calculations, in function \texttt{calc\_optics} (see Figure \ref{fig:diagram}). Specifically, our final goal is to obtain species-volume-integrated layer optical depth, single scattering albedo, and asymmetry as a function of both pressure layer and wavelength. This relies on our pre-computed Mie grid, which has stored: the extinction efficiency (Q$_\mathrm{ext}$), scattering efficiency (Q$_\mathrm{scat}$), and asymmetry parameter multiplied by the scattering efficiency ($g*$Q$_\mathrm{scat}$). We 
describe this calculation in detail in \S \ref{sec:mie}, which users can either do themselves or download the default data \citep{virga_iors_1}. \footnote{\href{https://zenodo.org/records/15886530}{Zenodo}}.

To compute the volume-integrated optical properties across all species, $s$, we first define the probability density function, given the particle radius grid, $r$, the mean geometric particle radius, $r_g(s,z)$ (note we drop the functional terms $s$ and $z$ in the equations below), and particle width distribution, $\sigma_g$: 
\begin{equation}\label{eqn:PDF}
    \textrm{PDF}_r(r;r_g,\sigma_g) = \frac{1}{r \sqrt{2\pi} \ln{\sigma_g}} \exp \left({- \frac{\ln^2(r/r_g)}{2\ln^2\sigma_g}} \right)
\end{equation}
where the lognormal size distribution per unit radius interval, $\Delta r'$ , is given by 
\begin{equation}
    \frac{dN}{dr}(r|z,s) = \frac{N_{dz}(s,z)  \textrm{PDF}_r}{\sum_{r'} \textrm{PDF}_r(r';r_g,\sigma_g) \Delta r'}.
\end{equation}
$N_{dz}(s,z)$ is computed via Eqn. \ref{eqn:ndz}, and the sum over all radii in the denominator ensures our integral sums to 1. We can compute the volume Mie coefficients: 
\begin{equation} \label{eqn:fullparticledistribution}
    \tau_{\mathrm{ext}}(\lambda,z) = \sum_{r',s} Q_{\mathrm{ext}}(\lambda,r',s) \frac{dN}{dr}(r|z,s) \pi r'^2 \Delta r'
\end{equation}
where we similarly compute Q$_{\mathrm{scat}}(\lambda,z)$ and $g(\lambda,z)$. Finally, we compute single scattering albedo ($w_0(\lambda,z)$=Q$_{\mathrm{scat}}$/Q$_{\mathrm{ext}}$), asymmetry parameter ($g_0(\lambda,z)=g$), and the extinction optical depth (opd$(\lambda,z)$=$\tau_{\mathrm{ext}}$). These latter three properties are ultimately what is passed to radiative transfer solvers such as \texttt{PICASO}. Note that \texttt{PICASO} is directly coupled to \texttt{Virga} and the two can be easily used together\footnote{\href{https://github.com/natashabatalha/virga/blob/d543c388f9bd8603b8a3c94a0ae09f9befadec40/docs/notebooks/4_PairingOutputToPICASO.ipynb}{PICASO-Virga coupling tutorials}}.  

\section{Modeling Details} \label{sec:details}

\subsection{Optical Constants Data}

\begin{figure*}
    \centering
    \includegraphics[width=0.7\linewidth]{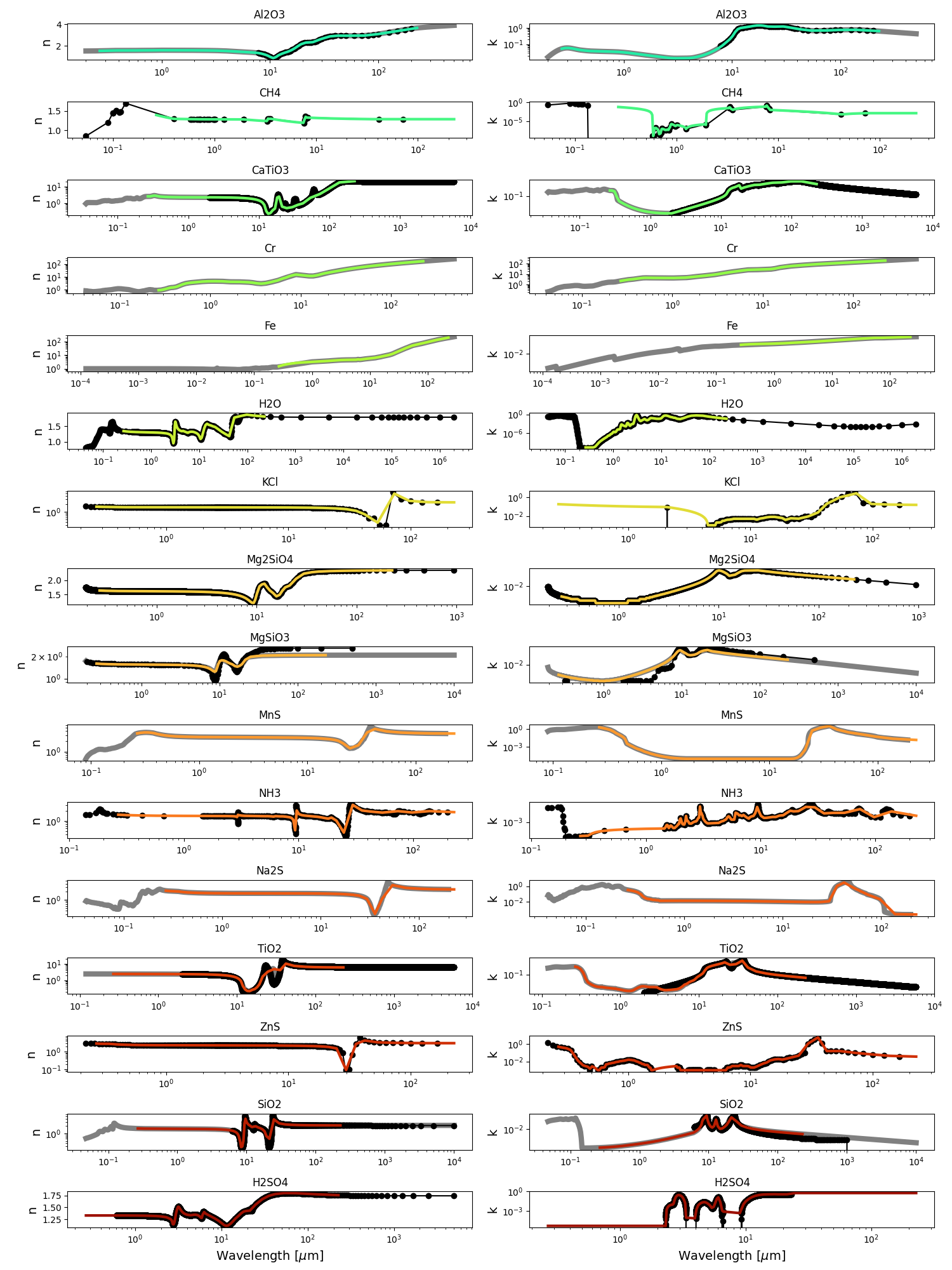}
    \caption{Processed index of refraction for the condensates available in \texttt{Virga}.  Grey lines represent data available through the HITRAN2020 \citep{HITRAN2020} aerosol database. Black circles represent data available from the LX-MIE database \citep{Kitzmann2018optical}. Colored lines represent the interpolated and processed index of refraction from which our Mie properties are computed. This figure can be fully reproduced and modified with new data using \texttt{Virga}'s IOR Factory \href{https://github.com/natashabatalha/virga/blob/d543c388f9bd8603b8a3c94a0ae09f9befadec40/docs/notebooks/5_IndexofRefractionFactory.ipynb}{(link here)}.  }
    \label{fig:ior}
\end{figure*}

Optical constants are made up of the wavelength-dependent real, $n$, and imaginary, $k$, part of the refractive index. The real component is related to the scattering properties of particles, while the imaginary component describes their absorptive properties. These constants are usually derived from disparate experimental measurements that are not necessarily specific to exoplanet atmospheric conditions. In many cases measurements from a single data source cover a narrow wavelength range at a single pressure and temperature. Therefore, to obtain a ``complete'' data set, sources often must be aggregated and stitched together. 

We define a fully ``complete'' dataset for a single condensate as one that has both $n$ and $k$ for the wavelength range $\sim$0.3-300$\mu$m. This wavelength range is necessary to study the radiative effect that clouds have on the atmosphere, and/or to study the impact of clouds on the observable spectra of exoplanets and brown dwarfs (e.g. with \textit{JWST} from 0.7-28$\mu$m). For the purposes of studying radiative effects, only a low resolution wavelength grid is needed. For example, the grid originally used in AM01  is a 196-point wavelength grid with $\delta\lambda\sim$0.008-88~$\mu$m, corresponding to R$\sim$2.5-250. It was adopted from the grid used in the one-dimensional climate model developed to study the atmosphere of Titan \citep{mckay1989thermal}, Uranus \citep{marley1999thermal}, brown dwarfs \citep{marley2002clouds}, and later, exoplanets \citep{fortney2008unified}. For studies of cloud features in spectra (e.g., with JWST) higher resolution wavelength grids are needed. For example, the detection of quartz clouds in WASP-17b, which used the \texttt{Virga} code base \citep{Grant+2023}, required a R=300 wavelength grid from 5-14$\mu$m. 

Several studies have compiled these data sources to obtain a complete set of optical constants. The original AM01 works focused on cloud properties of silicates, in the form of \ce{MgSiO3}, \ce{H2O}, and iron. Later, \citet{morley2012neglected}, expanded on that condensate dataset by adding KCl \& ZnS from \citet{querry1987optical}, MnS from \citet{Huffman1967MnS}, Cr from \citet{Stashchuk1984Cr} and \ce{Na2S} from \citet{Montaner1979Sulfur} and \citet{Khachai2009Na2S}. These enabled the computation of cloud models for exoplanets and brown dwarfs in the regime of T$_\mathrm{eff}$=400--1300~K. To fill the critical gap needed understand the cloud contributions seen in atmospheres of hot Jupiters \citep[e.g.][]{Sing2016continuum}, \citet{Wakeford2015clds} and \citet{Wakeford2017hightemp} compiled a set of optical constants for 20 more condensate species which covered T$_\mathrm{eff}$=70--1700~K. In this set, \ce{C6H12} and \ce{Al2O3} bookend the lowest and highest condensation temperatures, respectively. \citet{Kitzmann2018optical} revisited and expanded upon the dataset of \citet{Wakeford2015clds} by 1) adding new dust species, 2) updating data when necessary, and 3) using Kramers-Kronig relations to maintain a physical relationship between $n$ and $k$. \citet{HITRAN2020} (HITRAN 2020) published an exoplanet specific optical condensate data base, which includes many of the same species found in either \citet{Wakeford2017hightemp} and/or \citet{Kitzmann2018optical}.  Lastly, \citet{Luna2021clouds} updated a subset of cloud properties needed for brown dwarf and gas giant planets (namely Al-, Ca-, Fe-, Mg-, and Si- based species). \citet{Luna2021clouds} combine refractive indices from each crystallographic axis by computing absorption and scattering efficiencies first, then averaging each axes together. \citet{Kitzmann2018optical} took a similar approach, weighting the dielectric function along crystallographic axis by 1/3 before computing $n$ and $k$ for anisotropic materials.

In this work we do not aim to re-aggregate optical constant data. Instead, our goal is to provide a simple, open-source, and reproducible method to transform raw optical constant data source(s) to a single dataset per condensate that exists on a uniform wavelength grid. This is not trivial. For example, at the time of this publication, HITRAN 2020 includes one selected optical constant data source per condensate, making it difficult to obtain a complete-in-wavelength dataset. \citet{Kitzmann2018optical} uses multiple data sources but has post-processed from the raw data sources, which makes it difficult to add further data updates.  \citet{Wakeford2015clds} also publishes the post-processed and interpolated refractive and imaginary indices of refraction. \citet{Mullens2024} collated an extensive set of condensate optical properties from the above sources for POSEIDON, also using Kramers-Kronig relations to fill in gaps in data.

Included in Table \ref{tab:optical} is a full list of the condensate species included in \texttt{Virga}, the source of the raw laboratory- or theoretical- data and the source of the data aggregation. Our decision tree for choosing data is as follows: 
\begin{itemize}
    \item  Is a complete-in-wavelength data available through HITRAN 2020? 
    \subitem If yes, use it
    \subitem If no, is a complete-in-wavelength data available direct from online publication? 
    \subsubitem If yes, use it 
    \subsubitem If no, use the aggregated set from \citet{Kitzmann2018optical}
\end{itemize}
In this way we minimize the post-processing done from the original data source to the final product used by \texttt{Virga}. Users have the flexibility to create an Index of Refraction (IOR) database on custom wavelength-bins. The \texttt{Virga} ``Index of Refraction (IOR) factory'' source code can be on \texttt{Github}\footnote{ \href{https://github.com/natashabatalha/virga/blob/d543c388f9bd8603b8a3c94a0ae09f9befadec40/virga/ior_factory.py}{IOR factory source code}} along with the notebook tutorial, which shows how to create a custom database\footnote{\href{https://github.com/natashabatalha/virga/blob/d543c388f9bd8603b8a3c94a0ae09f9befadec40/docs/notebooks/5_IndexofRefractionFactory.ipynb}{Tutorial on creating custom-in-wavelength IOR databases} }. The final result of our interpolated index of refraction database, which is available on Zenodo now as Version 2,  \citep{virga_iors_1} is shown in Figure~\ref{fig:ior}.

    
\begin{table*}[]
    \centering
    \begin{tabular}{c|c|c|c|c|c}
    \hline  \hline 
    \textbf{Condensate} & \textbf{Raw data source(s)} & \textbf{Wave. Range} & \textbf{Agg. by} & \textbf{Sampling} & \textbf{Comments } \\
     & & $\mu$m & & $\delta\mu$m & \\
    \hline
    \ce{Al2O3} & \citet{Koike1995Al2O3}   & 0.2--12 & DK18 & 0.008--19 & \\
          & \citet{Begemann1997Al2O3} & 7.8--500 &  & & amorphous at 873 K \\
    \ce{CaTiO3} & \citet{Posch2003TiOs} & 2-5.8e5 & DK18 & 0.0005--4 & \\
    & \citet{ueda1998CaTiO3} & 0.02-2 & & & \\ 
    \ce{CH4}(l) & \citet{Martonchik1994CH4} & 0.0545--71.43 &  & 0.001--33 & gaps/\texttt{numpy.interp} \\
     &  &  & &  & liquid at 90~K \\
    Cr & Lynch \& Hunter in \citet{palik1991Vol2} & 0.04--31 & DK18 & 0.0005--5 &  \\ 
       & \citet{Rakic1998Optical} & 31--500 &  & 1--15 &  \\   
    Fe & Lynch \& Hunter in \citet{palik1991Vol2} & 1e-4--285 &  & 3e-6--85 &  \\ 
    \ce{H2O} & \citet{warren2008H2O} & 0.0443--2e6 & HT20 & 0.0003--1e6 &  \\ 
    KCl & \citet{querry1987optical} & 0.22--166.67 & HT20 & 0.01--42 &  \\     
    \ce{Mg2SiO4} & \citet{jager2003mgs} & 0.20--948 & HT20 & 4e-5--474 & amorphous(sol-gel) \\ 
    \ce{MgSiO3} & \citet{jager2003mgs} & 0.2--500.0 & DK18 & 0.02--300 & amorphous(sol-gel) \\
    MnS & \citet{Huffman1967MnS} & 0.05-13 & DK18 & 0.0005--4 & \\
    & \citet{Montaner1979Sulfur} & 2.5 -- 200 & & & \\ 
    \ce{NH3} & \citet{Martonchik1984NH3} & 0.14--200 &  & 0.0003--33 & \\
    \ce{Na2S} & \citet{Khachai2009Na2S} & 0.04-63 & DK18 & 0.0005--4 & \\
    & \citet{Montaner1979Sulfur} & 2.5--200 & & & \\ 
     \ce{SiO2} & Philipp in \citet{Palik1985hocs.book.....P} & 0.05 -- 8.4 & DK18 & 0.0005--4 &  anisotropic \\
     & \citet{Zeidler2013SiO2} & 6.25 -- 10$^4$ &  & & $\alpha$-crystal at 928 K\\
    \ce{TiO2} & \citet{Zeidler2011TiO2} & 0.4--10 & DK18 & 0.0005--4 & Anatase AB \\
    & \citet{Posch2003TiOs} & 10--6e3 & & & \\ 
    & \citet{Siefke2016TiO2} & 0.1--125 & & & \\ 
    \ce{ZnS} & \citet{querry1987optical} & 0.22-166.67 & HT20 & 0.01--42 & gaps/\texttt{numpy.interp} \\    
    \end{tabular}
    \caption{Optical property sources for condensates in \texttt{Virga}. DK18 \citet{Kitzmann2018optical}, HT20 \citet{HITRAN2020}}
    \label{tab:optical}
\end{table*}

\subsection{Calculating Mie Grid}\label{sec:mie}

Using the refractive indices described above, \texttt{Virga} includes the function \texttt{calc\_mie\_db} to generate a Mie efficiencies grid from these data. \texttt{Virga- v0} used the python package \texttt{PyMieScatt} \citep{sumlin2018} to compute Mie efficiencies, but this package is no longer actively developed. Therefore, we now use \texttt{MiePython v3.0.0} \citep{miepython} which has the built-in package for Mie grid computations. Given refractive indices $n$ and $k$ as a function of wavelength, \texttt{MiePython} outputs the extinction efficiency (Q$_\mathrm{ext}$), scattering efficiency (Q$_\mathrm{scat}$), and asymmetry parameter ($g$), which \texttt{Virga} then normalizes by the scattering efficiency ( $g$/Q$_\mathrm{scat}$). Within \texttt{calc\_mie\_db}, we compute Mie efficiency values for each radius in the Mie grid. 

To ensure both radiative and spectral effects of clouds are adequately captured, reasonable upper and lower particle radius bounds must be included when computing Mie efficiencies. The standard Mie grid that \texttt{Virga} uses ranges from 1$\times10^{-8}$ cm to 0.05 cm in 60 logarithmic steps. We consider at least 40 radii grid points to be the safe ``minimum'' number of radii to compute should a user choose to make a custom Mie grid. Mie coefficients produce interference fringes that occur due to the resonance for the scattering of a single particle \citep{Hansen1974}, so we include a sub-bin smoothing step in our calculation. Given the minimum and maximum radii, and the number of radii, the function \texttt{get\_r\_grid} defines bin centers, bin widths ($dr$), bin minima and bin maxima. Each bin is then divided into six further sub-bins defined by $dr$, linearly centered about each mean radius in the grid. Mie efficiencies are computed for each of the six sub-bin radii values, which are then averaged together to find the ultimate Mie efficiency values used for each radius value in the grid. 

When \texttt{Virga} computes a mean particle radius, it searches for values between 10$^{-10}$ -- 10 cm. Then, it computes the full particle distribution based on Eqn .\ref{eqn:PDF} ($r$ in this equation is the user pre-defined input Mie grid). If the code cannot find a particle radius solution within 10$^{-10}$ -- 10 cm it will not allow the calculation to proceed, and an error is raised. However, if a mean particle radius is found but the user-input $1\sigma$-distribution width exceeds the bounds of the user's Mie grid, the user will only receive a warning. In \texttt{v0} we experimented with having a full error returned to the user in this scenario. However, the error was too restrictive and we found most cases where this does occur to be when optical depths are low and do not influence spectra morphology or climate.  However, when users see this warning, which states: ``Take caution in analyzing results. There have been a calculated particle radii off the Mie grid, which has a min radius of X~cm and distribution of Y'', they should carefully evaluate whether to use the solutions found in these cases.  

\begin{table*}[]
    \centering
    \begin{tabular}{c|c|c|c|l|l}
    \hline  \hline 
         \textbf{Formula} & \textbf{Limiting} & \textbf{T cond (K)} & \textbf{Mixing} & \textbf{Saturation vapor}  & \textbf{References}  \\
          & \textbf{element} & \textbf{@ 0.1 bar}\footnote{\href{https://natashabatalha.github.io/virga/notebooks/1_GettingStarted.html\#How-to-Compute-Temperature-Condensation-Curves}{Tutorial: How to get condensation temperatures with \texttt{Virga}}} & \textbf{ratio(ppm)}\footnote{\href{https://github.com/natashabatalha/virga/blob/d543c388f9bd8603b8a3c94a0ae09f9befadec40/virga/gas_properties.py}{Source code for all gas properties can be found here}; tabular abundances  for $\sim$ solar-composition gas.}  & \textbf{pressure (base-10 log bar)}\footnote{\href{https://github.com/natashabatalha/virga/blob/d543c388f9bd8603b8a3c94a0ae09f9befadec40/virga/pvaps.py}{Source code for all saturation vapor curves}} &  \\
    \hline
         CH$_4$(s) & \multirow{2}{*}{C} &  \multirow{2}{*}{57} & \multirow{2}{*}{490} & $\log p_{\textrm{CH}_4}'\approx 4.283 - 475.6/T;<90.6$~K & \multirow{2}{*}{Lo98} \\
         CH$_4$(l) & & & & $\log p_{\textrm{CH}_4}'\approx 4.092 - 459.8/T;>90.6$~K & \\
         NH$_3$(s) & \multirow{2}{*}{N} & \multirow{2}{*}{130} & \multirow{2}{*}{134} & $\log p_{\textrm{NH}_3}'\approx 6.90 - 1588/T;<195.4$~K & \multirow{2}{*}{Lo98} \\
         NH$_3$(l) & & & & $\log p_{\textrm{NH}_3}'\approx 5.201 - 1248/T;>195.4$~K & \\
         H$_2$O(s) & \multirow{2}{*}{O} & \multirow{2}{*}{230} & \multirow{2}{*}{754} & $\log p_{{\rm H_2O}}'\approx 7.610 - 2681/T; < 273.1$~K & \multirow{2}{*}{Bu81, Fl92, Lo98} \\
         H$_2$O(l) & & & & $\log p_{{\rm H_2O}}'\approx 6.079 - 2261/T; > 273.1$~K &  \\
         KCl & K & 749 & 0.255 & $\log p_{\rm KCl}'\approx 7.61 - 11382/T$ & Mo12, cf.~Lo99\\
         ZnS & Zn & 762 & 0.076 & $\log p_{\rm Zn}'\approx 12.81 - 15873/T - [\textrm{M/H}]$ & Vi06, Mo12  \\
         Na$_2$S & Na & 932 & 3.34 & $\log p_{\rm Na}'\approx 8.55 - 13889/T - 0.5[\textrm{M/H}]$ & Vi06, Mo12\\
         MnS & Mn & 1275 & 0.541 & $\log p_{\rm Mn}'\approx 11.53-23810/T - $[M/H] & Vi06, Mo12\\
         Cr & Cr & 1426 & 0.887 & $\log p_{\rm Cr}'\approx 7.49 - 20592/T$ & Mo12 \\
         Mg$_2$SiO$_4$\footnote{The cloud base mixing ratio of Mg$_2$SiO$_4$ and MgSiO$_3$ are highly dependent on whether or not both clouds are assumed to condense. Here, to remain consistent with \citet{diamondback} we assume both condense. The mixing ratio for Mg$_2$SiO$_4$ in \citet{Gao2020Aerosol} is roughly a factor of 2 higher because it only assumes formation of \ce{Mg2SiO4}.} & Mg & 1515 & 30.63 & $\log p_{\rm Mg}' \approx 14.88 -32488/T - 0.20\log P - 1.4 $[M/H] & Mo24, cf.~Vi10\\
         MgSiO$_3$ & Si & 1603 & 29.20 & $\log p_{\rm SiO}'\approx 13.43 - 28665/T - $[M/H] & Mo24, cf.~Vi10\\
         CaTiO$_3$ & Ti & 1650 & 2.51 & $\log p_{\rm TiO}'\approx 30.24 - 72160/T - \log P - 3 $[M/H] & cf.~Wa17, Lo02 \\
         Fe & Fe & 1697 & 50.95 & $\log p_{\rm Fe}'\approx 7.09-20995/T$ & Mo24, cf.~Vi10 \\
         SiO$_{2}$ & Si & 1540 & 60.3 & $\log p_{\rm SiO}' \approx 13.168 - 28265/T - [{\rm M/H}]$ & Gr23 \\
         TiO$_{2}$ & Ti & 1879 & 0.169 & $\log p_{\rm TiO}' \approx 13.95 - 38266/T - [\rm M/H]$ & cf.~Ma21, Lo02 \\
         Al$_{2}$O$_{3}$ & Al & 1892 & 2.489 & $\log p_{\rm Al}' \approx 15.24 - 41481/T - 1.50[\textrm{M}/\textrm{H}]$ & Mo24, cf.~Wa17, Lo02 \\ \hline \hline
    \end{tabular}
    \caption{\textbf{Properties of the available condensates in \texttt{Virga}}. References: 
    Bu81: \citet{buck1981new}; 
    Lo98: \citet{Lodders1998book}; 
    Fl92: \citet{flatau1992polynomial};
    Lo02: \citet{Lodders2002titanium}    
    Vi06: \citet{Visscher2006atmospheric};     
    Lo10: \citet{Lodders2010SolarSystem}, 
    Vi10: \citet{Visscher2010atmospheric}; 
    Mo12: \citet{morley2012neglected};     
    Wa17: \citet{Wakeford2017hightemp}; 
    Ma21: \citet{Marley-sonora-2021ApJ...920...85M};  
    Gr23: \citet{Grant+2023};
    Mo24: \citet{diamondback}
    }
    \label{tab:pvap}
\end{table*}


\subsection{Condensate Formation Model}\label{sec:cloudcond}
The formation of condensates is calculated assuming vapor pressure saturation for cloud-forming species. This equilibrium condensation condition is met when the atmospheric partial pressure of a condensable gas species $j$ ($p_j$) becomes greater than the saturation vapor pressure of $j$ ($p_j'$):
\begin{equation}\label{equation: condensation condition}
    p_j \geq p_j',
\end{equation}
where $p_j$ is the atmospheric partial pressure of $j$ and $p_j'$ is the temperature-dependent vapor pressure of $j$ over the condensate. The values for $p_j$ for each condensable gas are calculated from pre-defined mixing ratio values stored in \texttt{virga.gas\_properties}:
\begin{equation}\label{equation: mixing ratio definition}
    p_j = X_jp_t,
\end{equation}
where $X_j$ is the expected volume mixing ratio of the condensable species $j$ at the cloud base and $p_t$ is the total pressure. The values of the solar-composition mixing ratios ($X_j$) defined in \texttt{virga.gas\_properties} are given in Table \ref{tab:pvap}. In Virga we make the assumption that volume mixing ratios varies linearly with M/H factor below the cloud deck. Note that users can overwrite these gas mixing ratios by using the key word argument, \texttt{gas\_mmr}, ( e.g., \texttt{gas\_mmr=\{`H2O':1e6\}} ). This allows users to pass their own chemistry calculations to the \texttt{Virga} code. Also note that for some species the volume mixing ratio $X_j$ may show non-linear dependence with metallicity and/or element gas-phase speciation, depending on the elemental composition and chemical behavior of species $j$ and the $p-T$ conditions of the atmosphere. 

The temperature-dependent values for the saturation vapor pressure ($p_{j}'$) are taken from \texttt{virga.pvaps} and listed in Table \ref{tab:pvap}. Note, that these are \textit{not} the same as ``condensation curves'' ubiquitously shown alongside P-T profiles, which users can derive using these vapor pressure curves in the function \texttt{virga.justdoit.condensation\_t}. The vapor pressure curves themselves are derived by first considering the net thermochemical reaction that produces the condensate from the most abundant gas-phase species of each of its constituent elements. In the next step, the condensable gas adopted for the vapor pressure calculation (and the gas species provided in \texttt{virga.gas\_properties} and listed in Table \ref{tab:pvap}) represents the most abundant gas-phase species of the limiting element for condensate formation, i.e.:
\begin{quote}
     if element $A$ is the limiting element of condensate $AB$, and gas $j(A)$ is the most abundant $A$-bearing gas, then the vapor pressure of $j(A)$ is used to evaluate Eq.~\ref{equation: condensation condition} for the condensation of $AB$
\end{quote}

As a brief illustration of this approach, we consider the formation of perovskite (CaTiO$_3$; e.g., \citealt{Wakeford2017hightemp}) via the net thermochemical reaction: 
\begin{equation}\label{reaction: perovskite-condensation}
    \rm Ca(g) + TiO(g) + 2H_2O(g) = CaTiO_3(c) + 2H_2(g),
\end{equation}
where Ca(g) is the most abundant Ca-bearing gas, TiO(g) is the most abundant Ti-bearing gas, and H$_2$O is the most abundant O-bearing gas near the expected condensation temperature. The Gibbs free energy function and the cloud base abundances of each vapor species in reaction (\ref{reaction: perovskite-condensation}) are used to solve for the saturation condition for perovskite. 
This is done by setting $p_j=p_j'$ and considering the Gibbs free energy function and the cloud base abundances of each vapor species in the reaction equilibria, e.g.,:
\begin{multline}\label{equation: perovskite-equilibria}
    \log p_{\rm TiO}' = \log (mX_{\rm TiO}p_t) \\ 
    \approx \frac{\Delta G_{\ref{reaction: perovskite-condensation}}(T)}{2.303RT} +2\log (X_{\rm H_2}p_t) \\
    - 2\log (mX_{\rm H_2O}p_t)-  \log (mX_{\rm Ca}p_t),
\end{multline}
where $m$ is the metallicity factor ($m=10^{[\rm Fe/H]}$). 
Because Ti is the limiting element in the production of CaTiO$_3$ (Ti$<$Ca$<$O in a solar-composition gas), the cloud base mixing ratio of TiO ($p_{\rm TiO}$) and the vapor pressure of TiO above CaTiO$_3$ ($p_{\rm TiO}'$) are used to evaluate Eq. \ref{equation: condensation condition}. 

For condensates produced by multiple gases from multiple elements (such as CaTiO$_3$), additional metal-bearing species (e.g., TiO, Ca, H$_2$O) maintain equilibrium with the condensate. In these cases, additional pressure and metallicity terms may be present in the vapor pressure approximation, because variations in their abundances may influence the saturation condition (Eq. \ref{equation: condensation condition}). For simpler systems (e.g., Fe, Cr, KCl, H$_2$O, NH$_3$, CH$_4$) the vapor pressure depends only upon a single condensable gas and the vapor pressure expressions ($p_j'$) are independent of metallicity. However, because cloud base mixing ratios ($X_j$) generally increase with [M/H], higher metallicities will tend to increase the temperature (i.e., depth) at which $p_j=p_j'$ and condensation occurs. Above the cloud base, the abundance of condensable gas $j$ is set by the vapor pressure $p_j'$.





%

\subsection{Eddy Diffusion Coefficient, K$_{zz}$ and Mixing Length}\label{sec:kzz}

There are three avenues for supplying K$_\textrm{zz}$ to \texttt{Virga}: 1) a pre-computed altitude-dependent profile, 2) a constant value, 3) the functional form of \citet{Gierasch1985}, which is computed via the function \texttt{get\_kz\_mixl}.

\citet{Gierasch1985} define the eddy diffusion coefficient as 
\begin{equation}
    K_{zz} = \frac{H}{3} \left( \frac{L}{H} \right)^{4/3} \left( \frac{RF_\textrm{conv}}{\mu \rho_a c_p} \right)^{1/3}
\end{equation}
where $H$ is the scale height ($=RT/\mu g$), $L$ is the turbulent mixing length, $R$ is the universal gas constant, $F_\textrm{conv}$ is the convective heat flux, $\mu$ is the atmospheric molecular weight, $\rho_a$ is the atmospheric density, and $c_p$ is the specific heat of the atmosphere. The convective heat flux, $F_\textrm{conv}$, is generally computed from layer-by-layer net fluxes in radiative-convective equilibrium models (e.g., \texttt{PICASO} or EGP) as: 
\begin{equation}
    F_\textrm{conv} = \sigma T_\mathrm{Teff}^4 - F_\mathrm{rad}
\end{equation}
where $F_\mathrm{rad}$ is the net upwards radiative flux. Convective overshoot (default $c$=\texttt{convective\_overshoot=None}) can be considered by ensuring that $F_\textrm{conv}$ does not exceed $c P_z/P_{z+1}$, where constant $c$ has typically been set to 1/3 (e.g., \citet{diamondback}). In cases where AM01 is used in isolation of radiative-convective equilibrium codes, a simplifying assumption that $F_\mathrm{net} = 0$ or that all the interior heat was transported through the convective heat flux can be used. Generally, though, $F_\textrm{conv}$ must be supplied through an external model. 

The mixing length parameter, $L$, is also computed in \texttt{get\_kz\_mixl} and can be modified by the \texttt{bool}, \texttt{latent\_heat}. When latent heat is turned on the mixing length is reduced from the atmospheric pressure scale height, $H$ 
\begin{equation}
    L = 
    \begin{cases}
        H \mathrm{max} (\Lambda , \Gamma_i/\Gamma_\mathrm{adiab}), & \text{latent\_heat=True} \\
        H, & \text{otherwise, default}
        \end{cases}
\end{equation}

where $\Gamma_i$ is the lapse rate of the layer and $\Gamma_\mathrm{adiab}$ is the dry adiabatic lapse rate, and $\Lambda$ is a minimum scaling applied, which we assume to be 0.1. Note historically, when AM01 was integrated into the legacy code EGP \citep{marley1999thermal}, $\Gamma_\mathrm{adiab}$, was calculated from a table of pre-computed lapse rates pertinent to H$_2$/He mixtures. This function can be seen in the corresponding \texttt{PICASO} \citep{Mukherjee2023ApJ...942...71M} function \href{https://github.com/natashabatalha/picaso/blob/46ada974afc621a8af5645c7abc50d69e1d435f2/picaso/climate.py#L15}{\texttt{did\_grad\_cp}} that is based on EGP. In the standalone \texttt{Virga} code, it is instead estimated as $\Gamma_\mathrm{adiab}=T_i/(7/2)$, where $7/2$ is the approximate specific heat factor of a permanent diatomic gas (e.g., H$_2$ or N$_2$). 

The last parameter input that factors into the user-prescribed K$_\textrm{zz}$ is the minimum threshold, default = 10$^5$cm$^2$/s. The AM01 rationale for this minimum threshold was the effect of breaking buoyancy waves \citep{1981JGR....86.9707L}. Logistically though, without a minimum threshold computed particle radii become numerically too small (e.g., in some cases smaller than a Bohr radius). 

\subsection{Calculation of Fall Speeds}\label{sec:fallspeed}
The fall speed calculation is used within \texttt{calc\_qc} and controls the computed mean particle radii. Here we describe this calculation in further detail. 

The viscous drag force exerted on a spherical particle falling through a theoretically infinite medium can be described using a relationship between the drag coefficient and the particle Reynolds number. 
For small Reynolds numbers, Stokes Law applies; however, outside of this region there is no adequate theoretical form for the drag coefficient. 
Instead, it is necessary to empirically determine the relationship by correlating to experimental data. 
Many different expressions have been adopted in the literature and some are applicable only to a limited range of Reynolds numbers. 
Furthermore, it is very difficult, if not impossible, to conduct experiments that accurately describe the motion of a condensate particle in an exoplanet atmosphere, primarily due to the diversity of such environments and also because we lack sufficient data to explain the precise physical properties. 
We must also account for non-continuum effects when the particle Knudsen number is very large and therefore the surrounding atmosphere should not be treated as a fluid with non-slip boundary conditions. 
In this section we analyze the existing relationship between the drag coefficient and Reynolds number being implemented in \texttt{virga}, highlighting discontinuities in the calculation, discussing modifications to ensure smoothness and attempting to include non-continuum effects where necessary.

\subsubsection{Background}
An object falling through a gas or fluid due to gravity reaches its terminal velocity when there is a balance between gravitational and viscous forces, namely
\begin{equation}
    F_g = F_\eta,
    \label{eq:forcebalance}
\end{equation}
where $F_g$ denotes the gravitational force on the particle in the direction of motion and $F_\eta$ denotes the opposing force due to viscous drag.
The drag force $F_\eta$ for a spherical particle of radius $r$ traveling through an atmosphere of density $\rho_a$ at speed $v_f$ is defined to be
\begin{equation}
    F_\eta = \frac{C_d}{2} \pi \rho_a r^2 v_f^2,
    \label{eq:viscous}
\end{equation}
where $C_d$ is the drag coefficient. 
The discrepancy between models that describe the terminal velocity for spherical particles in a fluid depends on the definition of $C_d$. 

\subsubsection{Existing terminal velocity calculation}
\label{sec:existing_calc}
The existing calculation for the terminal velocity implemented in \texttt{virga} involves the consideration of different Reynolds regimes to allow for different treatments of the drag coefficient $C_d$. 
We refer the reader to AM01 for a detailed description and summarize the calculation as follows:
\paragraph{Regime 1: $\Rey<1$}
For small Reynolds numbers $\Rey$, we assume Stokes flow \cite{stokes1851effect} corrected for non-continuum behaviour and define the fall speed for a spherical particle of radius $r$ and density $\rho_p$ as
\begin{equation}
    v_f = \frac{2\Cs(\rho_p-\rho_a) g r^2}{9\eta},
    \label{eq:stokes_vf_corrected}
\end{equation}
where $\rho_a$ and $\eta$ are the density and dynamic viscosity of the atmosphere respectively and $g$ is the acceleration due to gravity. The viscosity of the atmosphere in \texttt{virga} is computed in \texttt{layer} using the formalism outlined in \citet{rosner2012transport}: 
\begin{equation}
    \eta = \frac{5}{16} \frac{\sqrt{\pi k_B T \mu/N_A}}{\pi d_m^2}\frac{1}{1.22 (T/\epsilon)^{-0.16}}
\end{equation}
where $d_m$ is the diameter of the main atmospheric molecule (e.g. 2.8$\times$10$^{-8}$ cm for H$_2$), $\epsilon$ is the depth of the Lennard-Jones potential well for the atmosphere (e.g. 59.7 K for H$_2$), $\mu$ is the mean molecular weight, $T$ is the temperature of the layer. 

The parameter $\Cs$ in Eqn. \ref{eq:stokes_vf_corrected} is the Cunningham correction factor for non-continuum effects defined in \cite{cunningham1910velocity} to be
\begin{equation}
    \Cs = 1 + 1.26\,\Kn.
    \label{eq:slip_factor}
\end{equation}
for particle Knudsen number $\Kn$.

\paragraph{Regime 2: $1\leq\Rey\leq 1000$}
For intermediate Reynolds numbers, we note that the Davies or Best number $C_d\Rey^2$ \cite{pruppacher1980microphysics} is independent of velocity,
\begin{equation}
    C_d\Rey^2 = \frac{32 r^3 (\rho_p-\rho_a)\rho_ag}{3\eta^2}.
    \label{eq:davies_best}
\end{equation}
We fit $y=\log(\Rey)$ as a function of ${x=\log(C_d\Rey^2)}$ to data for rigid spheres from Table 10-1 of \cite{pruppacher1980microphysics}. 
The fit is given as a $6^\text{th}$ order polynomial, from which we calculate $\Rey$ and subsequently $v_f$ using
\begin{equation}
    \Rey = \frac{2r\rho_av_f}{\eta}.
    \label{eq:reynolds}
\end{equation}

\paragraph{Regime 3: $\Rey >1000$}
For high Reynolds numbers, we assume the drag coefficient is fixed at its asymptotic value $C_d=0.45$ \cite{pruppacher1980microphysics}, which by rearranging \eqref{eq:davies_best} and including the slip-correction factor $\Cs$ yields
\begin{equation}    
    v_f = \Cs\sqrt{\frac{8gr(\rho_p-\rho_a)}{3C_d\rho_a}}.
    \label{eq:vfall_regime3}
\end{equation}

\subsubsection{Fall speed algorithm}
\label{sec:algorithm}
To calculate the fall speed for a particle of radius $r$, \texttt{virga} currently implements the following algorithm:
\begin{enumerate}
    \item Assume $\Rey< 1$ (Regime 1) and calculate $v_f$ according to \eqref{eq:stokes_vf_corrected}.
    \item Recalculate $\Rey$ using \eqref{eq:reynolds} with $v_f$ derived in Step 1. 
        If $\Rey< 1$, output $v_f$. 
        If $1\leq\Rey\leq 1000$, go to Step 3.
        If $\Rey>1000$ go to Step 4.
    \item Calculate $v_f$ using the methodology outlined in Regime 2.
    \item Calculate $v_f$ using the methodology outlined in Regime 3, namely \eqref{eq:vfall_regime3}.
\end{enumerate}

\subsubsection{Non-continuum effects}
\begin{figure}[t!]
    \centering
    \begin{subfigure}[b]{0.48\textwidth}   
        \centering
        \includegraphics[width=\linewidth]{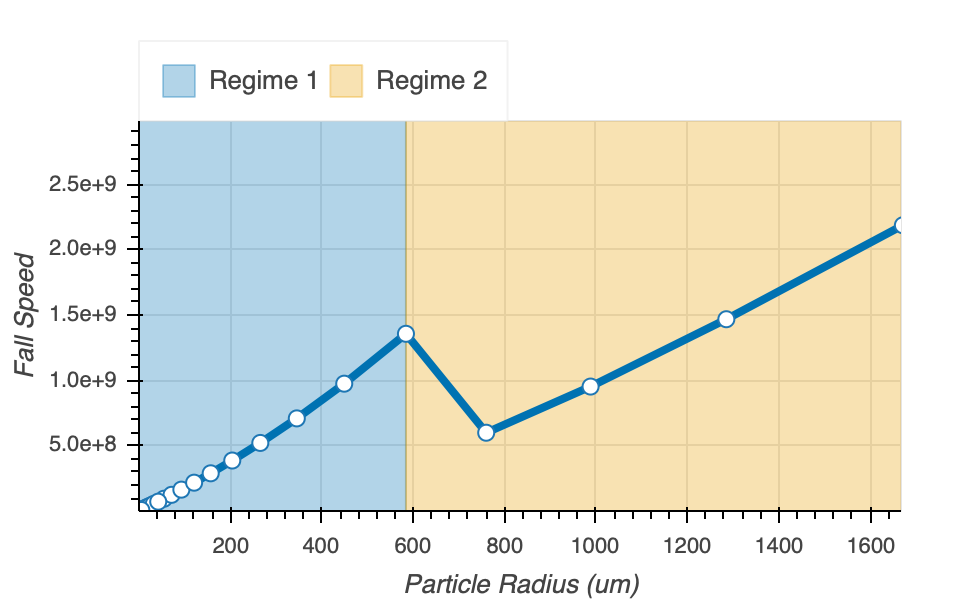}
        \caption{}\label{fig:1e-4_disc_radius}
    \end{subfigure}
    \begin{subfigure}[b]{0.48\textwidth}   
        \centering
        \includegraphics[width=\linewidth]{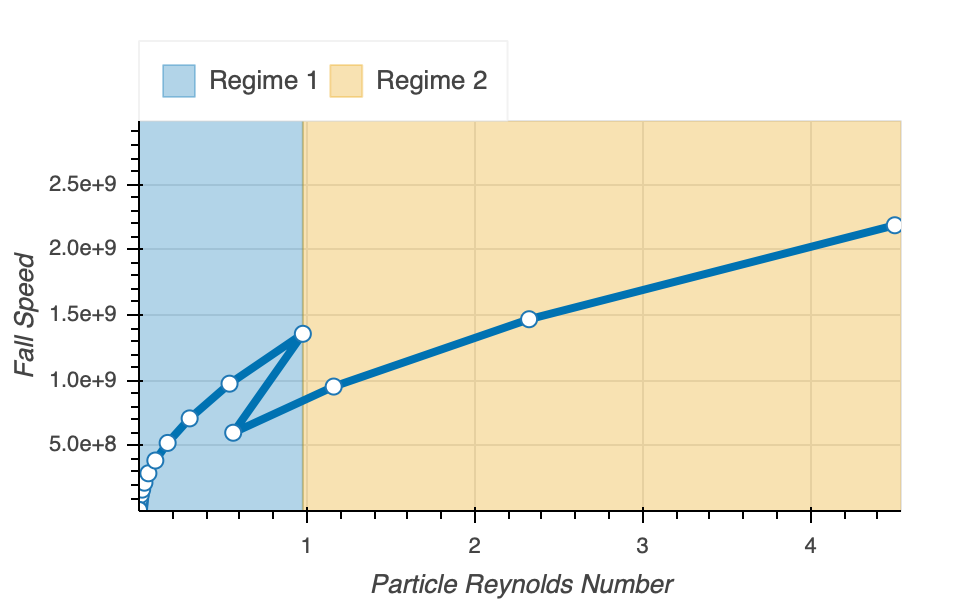}
        \caption{}\label{fig:1e-4_disc_reynolds}
    \end{subfigure}
    \caption{Fall speed for Cr particles at a pressure of $10^{-4}$ bars.}
    \label{fig:1e-4_disc}
\end{figure}
In Figure \ref{fig:1e-4_disc} we plot the terminal velocity against both the particle radius and the particle Reynolds number to analyze the transitions between the Reynolds regimes above.
The fall speeds are calculated according to the algorithm above and plotted in Figure \ref{fig:1e-4_disc_radius}. 
We illustrate the regime in which the terminal velocity was calculated by colored blocks (blue for Regime 1, orange for Regime 2).
The Reynolds numbers are then recalculated using \eqref{eq:reynolds} with the fall speed $v_f$ outputted by the algorithm and plotted in Figure \ref{fig:1e-4_disc_reynolds}. 

We see from Figure \ref{fig:1e-4_disc_radius} that there is a discontinuity when moving from Regime 1 to Regime 2
The terminal velocity is considerably reduced for a radius of approximately 780$\mu$m compared with a particle of radius $600\mu$m.
By turning our attention to Figure \ref{fig:1e-4_disc_reynolds}, we notice similarly strange behavior of the Reynolds numbers at this transition. 

The issue is down to non-continuum effects for which we corrected the drag coefficient with $\Cs$ \eqref{eq:slip_factor} in Regime 1, yet neglected in Regime 2.
The derivation of drag coefficients from the Navier-Stokes equations often assume continuum flow by treating the surrounding atmosphere as a fluid, thus enforce a no-slip boundary condition between the particle and surrounding atmosphere \citep{batchelor2000introduction}.
This assumption is no longer valid for particles whose radius approaches the molecular mean free path of the atmosphere $\lambda_\mathrm{mfp}$, in particular particles whose Knudsen number $\Kn$ satisfies
\begin{equation}   
    \Kn = \frac{\lambda_\mathrm{mfp}}{r} > 0.1.
    \label{eq:kn>0.1}
\end{equation}
If the Knudsen number satisfies \eqref{eq:kn>0.1}, the surrounding atmosphere can no longer be treated as a fluid and the no-slip boundary condition between the particle and the atmosphere used to derive the drag coefficient is no longer valid.
We must therefore correct the drag coefficient with the slip-correction factor \eqref{eq:slip_factor}.

The correction of the drag coefficient for Stokesian flow is prevalent, however the treatment of non-Stokesian behavior is difficult to come by.
This is because the Knudsen number is large for small particles, and small particles typically have low Reynolds numbers and hence fall into the Stokesian regime.
However, the Knudsen number also depends on the molecular mean free path $\lambda_\mathrm{mfp}$ of the atmosphere which depends on the temperature-pressure profile.  
For exoplanets, $\lambda_\mathrm{mfp}$ can be very large, thereby increasing the Knudsen number for radii whose Reynolds numbers are outside of the Stokesian regime. 
\begin{figure}[t!]
    \centering
    \includegraphics[width=\linewidth]{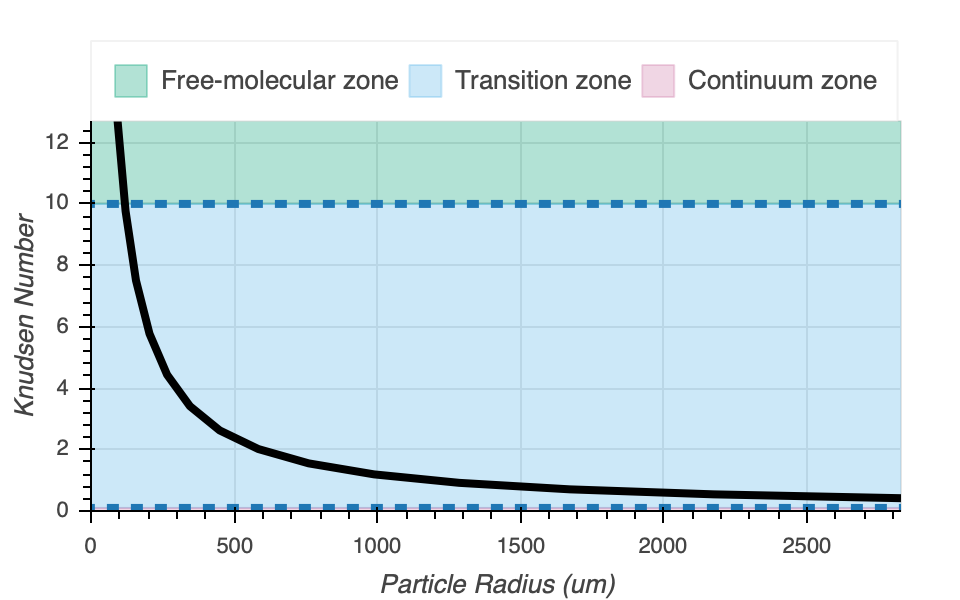}
    \caption{Knudsen numbers for Cr particles at a pressure of $10^{-4}$ bars.}
    \label{fig:1e-4_kn}
\end{figure}

We demonstrate this phenomenon in Figure \ref{fig:1e-4_kn}.
For every particle radii in our range, the Knudsen number $\Kn>0.1$, indicating that for every radius we are outside of the continuum zone, meaning we must include a slip-correction factor in our drag coefficient.

\begin{figure}[t!]
    \centering
    \begin{subfigure}[b]{0.48\textwidth}   
        \centering
        \includegraphics[width=\linewidth]{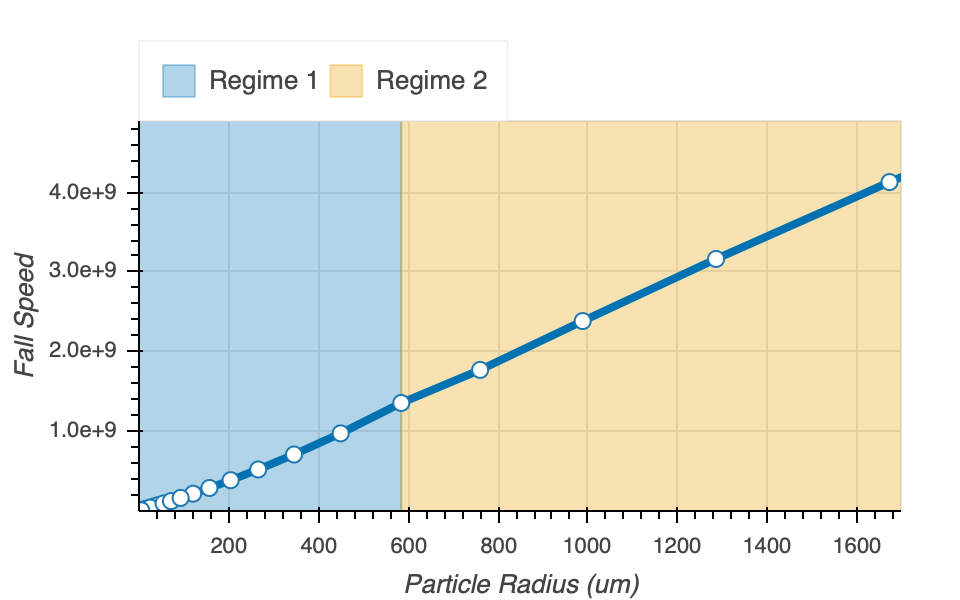}
        \caption{}\label{fig:1e-4_corrected_radius}
    \end{subfigure}
    \begin{subfigure}[b]{0.48\textwidth}   
        \centering
        \includegraphics[width=\linewidth]{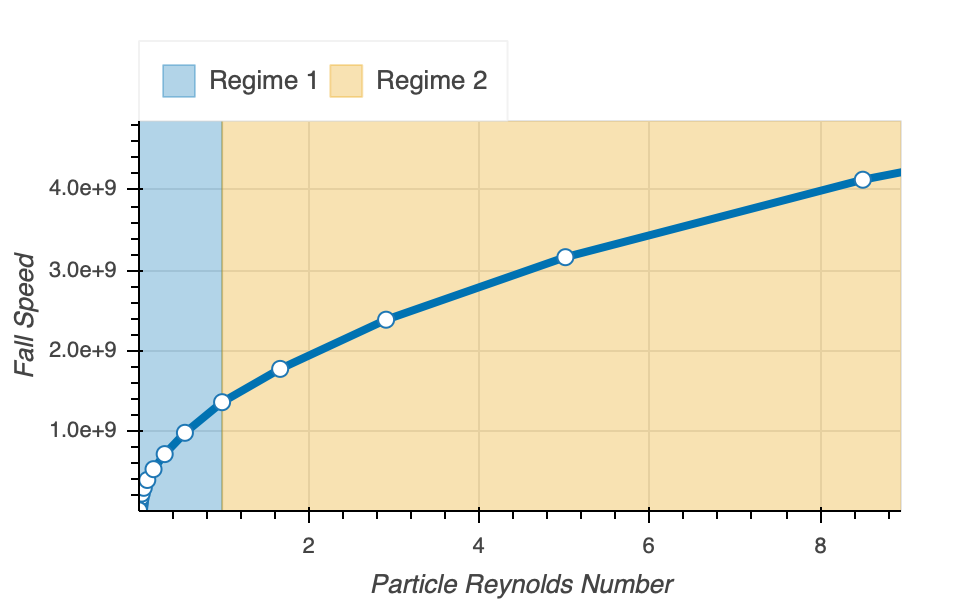}
        \caption{}\label{fig:1e-4_corrected_reynolds}
    \end{subfigure}
    \caption{Fall speed for Cr particles at a pressure of $10^{-4}$ bars following slip-correction in Regime 2.}
    \label{fig:1e-4_corrected}
\end{figure}

We modify our Regime 2 calculation in Section \ref{sec:existing_calc} by dividing the drag coefficient $C_d$ by the slip-correction factor $\Cs$ \eqref{eq:slip_factor}.  The resulting fall speeds and subsequent Reynolds numbers are plotted in Figure \ref{fig:1e-4_disc}. 
The results are plotted in Figure \ref{fig:1e-4_corrected}. 
We observe a smooth transition between Regimes 1 and 2, indicating that the drag coefficient in Regime 2 tends to Stokes drag as $\Rey\rightarrow 1$, as we would expect.

\subsubsection{Khan-Richardson model}
We can bypass the complexity of different regimes by considering the Khan-Richardson model \citep[henceforth referred to KR]{khan1987resistance}  to approximate the drag coefficient $C_d$. 
The KR drag is valid for  a large range of Reynolds numbers, $10^{-2}<N_\text{Re}<10^5$ \citep{richardson2002coulson} given by
\begin{equation}
    C_d(\Rey) = 2\left(1.849 N_\text{Re}^{-0.31} + 0.293 N_\text{Re}^{0.06}\right)^{3.45}.
    \label{eq:KR_drag}
\end{equation}
However, we have already mentioned that non-continuum effects are important for the environment we are considering, therefore we modify \eqref{eq:KR_drag} with the division of the Cunningham correction factor \eqref{eq:slip_factor} to obtain
\begin{equation}
    C_d(\Rey) = \frac{2}{\Cs}\left(1.849 N_\text{Re}^{-0.31} + 0.293 N_\text{Re}^{0.06}\right)^{3.45}.
    \label{eq:KR_drag_corrected}
\end{equation}
By considering the KR drag coefficient \eqref{eq:KR_drag_corrected} in the force balance \eqref{eq:forcebalance}, we can calculate the terminal velocity.

\subsubsection{Comparing terminal velocities}
\begin{figure}[b!]
    \centering
    \begin{subfigure}[b]{0.48\textwidth}   
        \centering
        \includegraphics[width=\linewidth]{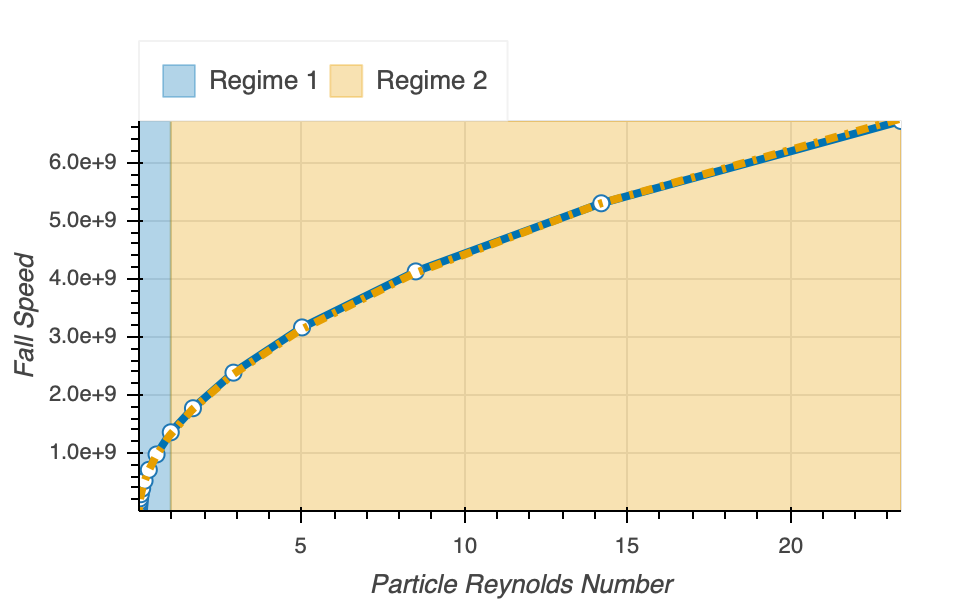}
        \caption{$10^{-4}$ bars}
        \label{fig:1e-4_compare_reynolds}
    \end{subfigure}
    \begin{subfigure}[b]{0.48\textwidth}   
        \centering
        \includegraphics[width=\linewidth]{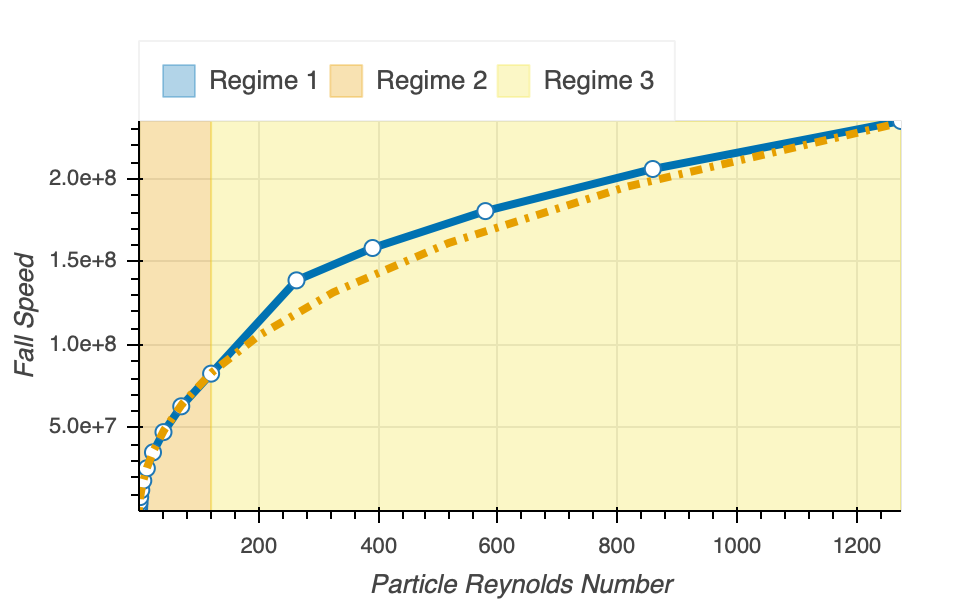}
        \caption{$10$ bars}
        \label{fig:1e0_compare_reynolds}
    \end{subfigure}
    \caption{Fall speed for Cr particles at different pressures obtained using the original regime-based calculation of \texttt{virga} (solid blue line) and the KR model (dashed orange line).}
    \label{fig:comparison}
\end{figure}
We consider two test pressures at which to compare the terminal velocity of Cr particles, namely $10^{-4}$ bars (Figure \ref{fig:1e-4_compare_reynolds}) and 10 bars (Figure \ref{fig:1e0_compare_reynolds}). 
We plot the terminal velocities versus Reynolds number for both the original calculation with the non-continuum correction in Regime 2 and the KR model.
Recall that the colored blocks highlight which regime was used to calculate the terminal velocity in the original calculation and the Reynolds number on the $x-$axis is obtained by considering this terminal velocity in \eqref{eq:reynolds}.

We notice that KR agrees with the regime-based calculation very well, in particular, the KR drag approaches Stokes drag for $\Rey<1$, as indicated by the strong agreement in Regime 1.
We also notice good agreement for large Reynolds numbers, however there is some disparity between the calculations for intermediate values.
This is to be expected and is a consequence of the different approximations in this region which, as we mentioned previously, are derived empirically by fitting to experimental data. 
Due to the lack of such data for an environment similar to that which we wish to model, there is no such approximation that accurately describes how a condensate particle falls.
Therefore, any model we choose here is subsequently limited and may not accurately reproduce the expected terminal velocity of a particle in an exoplanet atmosphere.

We note that the terminal velocity obtained using the original calculation in Regime 3 produces a Reynolds number of less than 1000, as illustrated in Figure \ref{fig:1e0_compare_reynolds}.
The is due to the terminal velocity algorithm outlined in Section \ref{sec:algorithm}.
In Step 2, we calculate the Reynolds number following the calculation of the terminal velocity in Step 1, thus defining our flow regime.
However, the terminal velocity in Step 1 was calculated under the assumption that $\Rey<1$. 
If $\Rey\geq1$ we have a contradiction of this assumption, and the terminal velocity we calculated is incorrect.
This is why we move out of the Stokesian Regime 1, however we cannot draw any conclusions about whether we are in Regime 2 or Regime 3 based off of the value of the Reynolds number.
Care must be taken following the contradiction of the assumption made in Step 1.


\subsubsection{Fall Speed Calculation Overview}
The original  terminal velocity calculation in AM01  contained  discontinuities between Reynolds regimes.
In \texttt{Virga} we removed the discontinuity between Regime 1 and Regime 2 by including the Cunningham correction factor \eqref{eq:slip_factor} in the drag coefficient Regime 2, demonstrating its necessity by studying the Knudsen numbers and proving the non-continuum nature of the environment.
We also point out a contradiction in Regime 3 where the terminal velocities computed in this region correspond to Reynolds numbers outside of the regime bounds.

\texttt{Virga} does contain an alternative calculation for the terminal velocity based upon the work of \cite{khan1987resistance}, triggered  when \texttt{og\_vfall=False}.
Though this is not the model default, it allows us to use a single expression for the drag coefficient, which we modified to be suitable for non-continuum behavior by dividing by the Cunningham correction factor, as is done for Stokesian Reynolds regimes.
This thereby results in a single expression for the terminal velocity, valid for all Reynolds regimes. 
This alternative approach bypasses any issues with continuity between different regimes, mitigating discontinuities and ensures smoothness of the terminal velocity with particle radius and Reynolds number.
This smoothness is particularly beneficial within numerical solvers and will be considered as the new default in future version of the code.

\subsubsection{Power-law approximation for determining $\alpha$}
\begin{figure}[t]
    \centering
    \includegraphics[width=\linewidth]{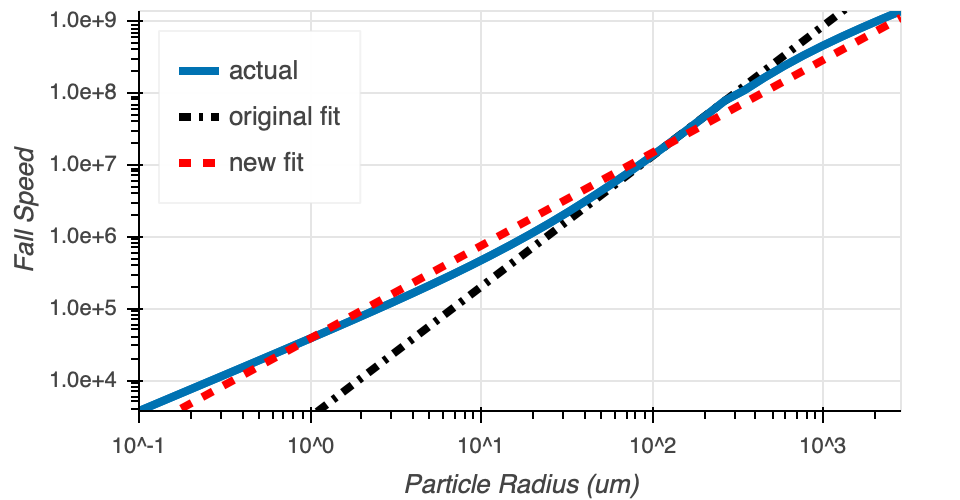}
    \caption{Power-law approximation to fall speed $v_f$ of Cr particles at 10$^{-2}$ bars. The new approximation (red dashed) is a better fit to the actual fall speed (blue solid) than the original calculation (black dash-dot).}
    \label{fig:power_law}
\end{figure}

AM01 fit a power-law approximation to the particle fall speed about its value at $v_f(r_w) = w_*$, that is when the fall speed is equal to the convective velocity $w_*$. The fit is given as
\begin{equation}
    v_f(r) = w_*\left(\frac{r}{r_w}\right)^\alpha,
\end{equation}
where the exponent $\alpha$ is calculated from a fit to $v_f$ at $r=rw/\sigma_g$ for $f_\text{sed}> 1$ and at $r=r_w\sigma_g$ otherwise, namely
\begin{align}
    \alpha=\begin{cases}
            \log(w_*/v_f)/\log\sigma, \quad f_\text{sed}> 1,\\
            \log(v_f/w_*)/\log\sigma, \quad f_\text{sed}\leq 1.
            \end{cases}
    \label{eq:alpha}
\end{align}
In \texttt{Virga} we now obtain a better fit by calculating $\alpha$ using \texttt{SciPy.optimize} function \texttt{curve\_fit} \citep{2020SciPy-NMeth}. 
We demonstrate the differences in the power-law approximations in Figure \ref{fig:power_law}.
We observe that the new fit with $\alpha$ calculated with \texttt{curve\_fit} is a better overall approximation of the true fall speed compared to the original fit using \eqref{eq:alpha}.

\begin{figure}[t!]
    \centering
    \includegraphics[width=\linewidth]{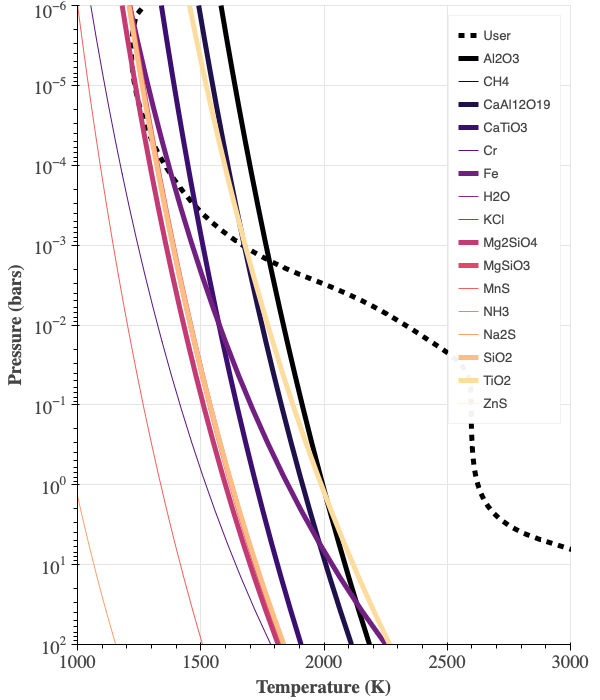}
    \caption{Default plot from \texttt{Virga}'s recommend gas function shown for the pressure-temperature profile published in \citet{Grant+2023}. Dashed curve represents the users input P-T profile, while solid lines represent the condensation curves. The bold solid curves are mathematically what gases \texttt{Virga} would condense. We urge users to carefully consider what condensate to include in their model instead of choosing all those that mathematically condense. }
    \label{fig:w17pt}
\end{figure}

\begin{figure*}[t!]
    \centering
    \includegraphics[width=\linewidth]{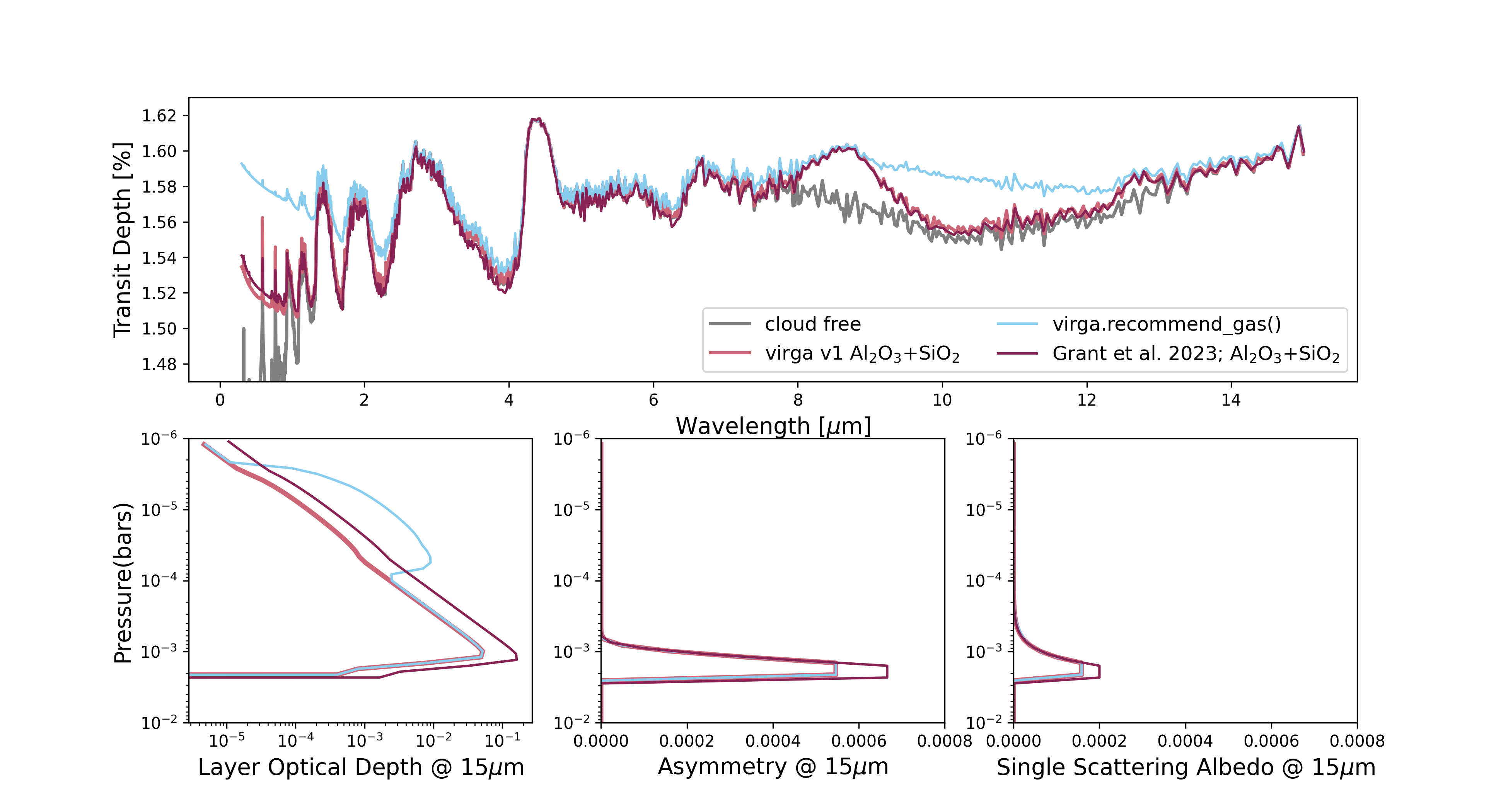}
    \caption{Here we demonstrate \texttt{Virga}'s ability to reproduce the results from \citet{Grant+2023}. The optical properties and spectra directly from \citet{Grant+2023} are shown in dark pink, which used \texttt{Virga-v0}. Light pink shows the results from \texttt{Virga-v1}, which differ slightly because of the updated Al$_2$O$_3$ condensate chemistry. Blue shows the model that would have been created had a user chosen all condensates available in \texttt{Virga-v1}. For reference, in grey, we show a cloud free spectrum to show where the spectral contribution from the cloud is dominant. }
    \label{fig:w17spec}
\end{figure*}

\section{Results: Exoplanet Benchmarks}\label{sec:benchE}

We illustrate the use of \texttt{Virga} for exoplanets by reproducing the  SiO$_2$ cloud detection in WASP-17 b \citep{Grant+2023}. We use the max loglikelihood pressure-temperature profile published in  \citet{Grant+2023} along with associated max loglikelihood model values for f$_\mathrm{sed}$ and  K$_\mathrm{zz}$. The \texttt{Virga} model published in \citet{Grant+2023} included Al$_2$O$_3$ and SiO$_2$ clouds only. However, these two gases are not the only two that condense along WASP-17 b's 1D pressure-temperature profile. We highlight this because \texttt{Virga} includes a function \texttt{recommend\_gas} that guides users to determine what condensates will mathematically condense, given the local pressure, temperature and chemistry assumptions, shown in Figure \ref{fig:w17pt}. Specifically Mg-clouds (MgSiO$_3$ and Mg$_2$SiO$_4$) are expected to condense at pressures just below SiO$_2$. However, the results of \citet{Grant+2023} demonstrate how users need to carefully consider what condensates to include in their model setup. 

Figure \ref{fig:w17spec} serves two purposes: 1) to show that we compare to the results from \citet{Grant+2023} using \texttt{Virga-v1}, and 2) to show the spectral difference between using all the condensates output by \texttt{Virga}'s recommend function and only the two presented in \citet{Grant+2023}. The optical depth values derived from \texttt{v1} are about a factor of 2 different from \citep{Grant+2023} which used \texttt{virga-v0}. This is because the gas condensate mixing ratio for Al2O3 was updated in \texttt{v1} from the value used in \citet{Gao2020Aerosol} (4.937~ppm) to be consistent with that published in \citet{diamondback} (2.489~ppm). This factor of 2 produces a minor effect on the spectrum but users should be aware of it anyways. 

Blindly using all the available condensates produces a second cloud deck at 0.1~mbar made of MgSiO$_3$ and Mg$_2$SiO$_4$. These condensates have a continuum-like spectral contribution from 10-12$\mu$m, which flattens out the water absorption feature detected by the MIRI LRS transit. We urge users to strongly consider their condensate set in the context of their substellar object, scientific use case, and any additional available observational context.

\section{Results: Brown Dwarf Benchmarks with Diamondback}\label{sec:benchB}

\begin{figure*}
    \centering
    \includegraphics[width=\linewidth]{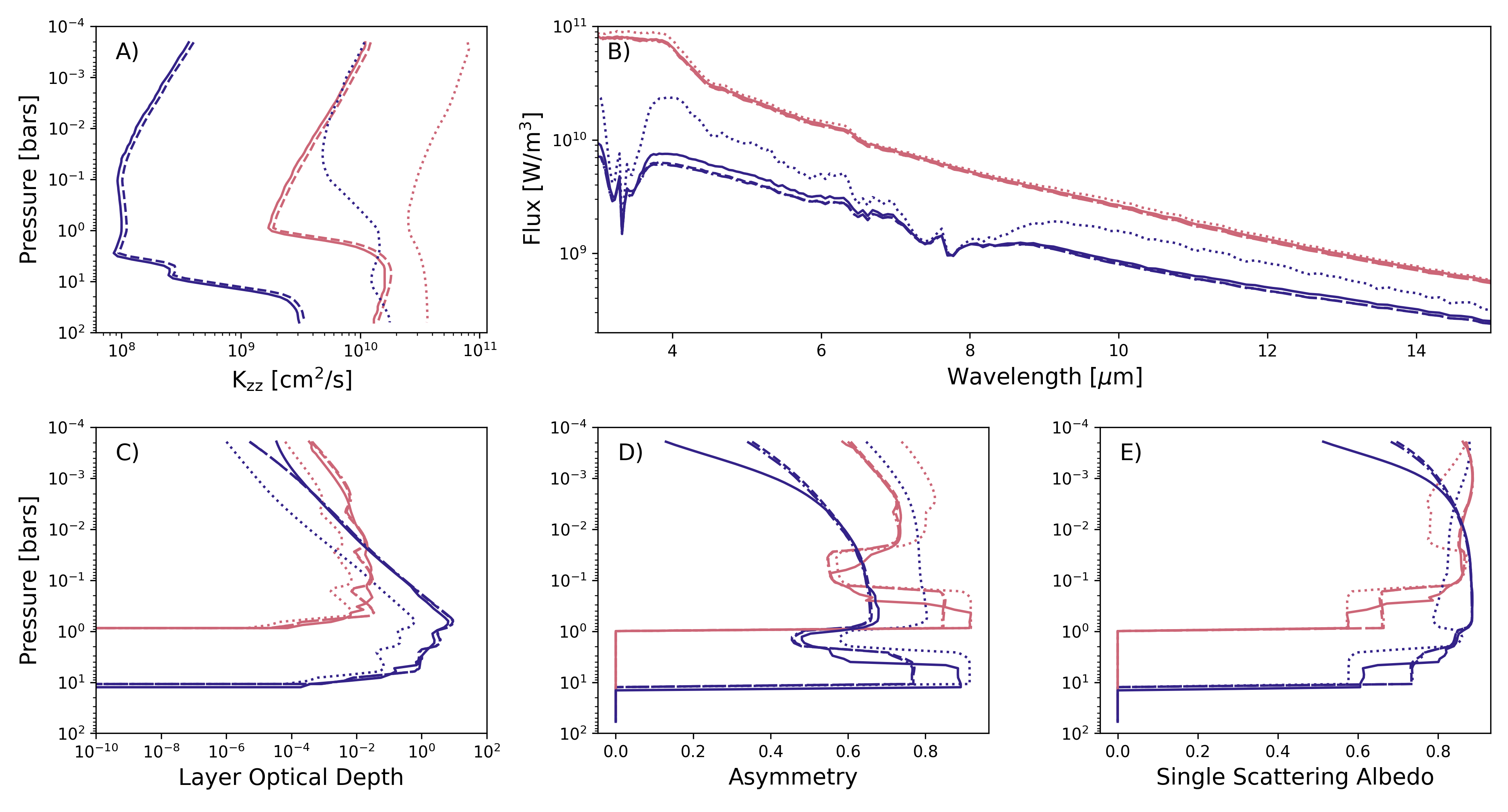}
    \caption{Here we demonstrate \texttt{Virga}'s ability to reproduce the Diamondback model grid's cloud models, given the pre-computed pressure-temperature profile. In A-E, solid lines are data derived straight from Zenodo or Diamondback log files. In A) we start by demonstrating the ability to reproduce the $K_{\rm{zz}}$ profile used for Diamondback using the \texttt{Virga}'s function to compute $K_{\rm{zz}}$. Dashed lines show the profile derived from a convective heat flux profile taken from EGP, while dotted lines shows the profile computed using an estimate for the convective heat flux (chf)  (dotted). Figures C-E show the  \texttt{Virga} output using the corresponding $K_{\rm{zz}}$ profiles from A). Note that dot-dash shows the optical properties derived when directly using the $K_{\rm{zz}}$ profile from Diamondback (solid line $K_{\rm{zz}}$ in A). B) shows all corresponding spectra alongside those published. Ultimately we can reproduce the Diamondback spectra when using self-consistent estimates for the convective heat flux. }
    \label{fig:diamondback1}
\end{figure*}

\begin{figure}
    \centering
    \includegraphics[width=\linewidth]{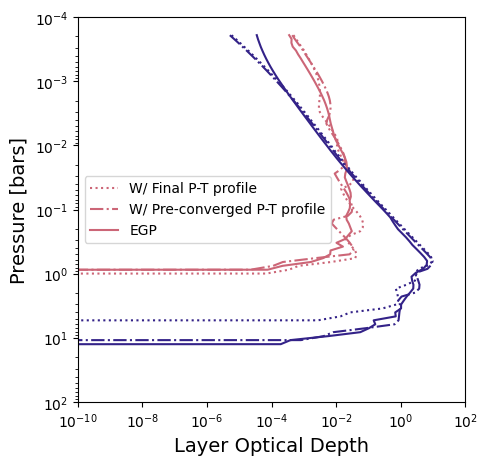}
    \caption{In some cases, the published pressure-temperature profiles from the Diamondback Zenodo repository \citep{morley_2024_12735103} (dotted) will produce slightly different optical depth profiles when reproduced with \texttt{Virga}. Here we show the reported Diamondback optical depth profiles available on Zenodo (solid), versus two \texttt{Virga} runs. The first is using the final converged pressure-temperature profile also published in the Diamondback Zenodo repository (dotted). The second is using the pre-converged pressure-temperate profile that was used for the final cloud run in EGP (dot-dash). In some cases these will agree  well (e.g. T$_{\mathrm{eff}}$=1900~K (pink)). In other cases users will notice slight discrepancies when attempting to reproduce Diamondback cloud models (e.g. T$_{\mathrm{eff}}$=900~K case). }
    \label{fig:diamondback2}
\end{figure}
We illustrate the use of \texttt{Virga} for brown dwarfs by showing examples reproducing Diamondback \citep{diamondback} model results for $T_\text{eff} = 900$K, $1900$K with $f_\text{sed} = 1$ and solar M/H and C/O. When reproducing self-consistent converged cloudy profiles, there is a subtlety one must consider. As the temperature profile iterates to convergence, both the $K_{\rm{zz}}$ and cloud profile are adjusted. The cloud profile relies on both the pressure-temperature profile and the $K_{\rm{zz}}$ profile. The $K_{\rm{zz}}$ profile relies on net fluxes, which also rely on the pressure-temperature profile. Convergence depends on the net fluxes, which depend on pressure-temperature, $K_{\rm{zz}}$, and the cloud profile. This circular problem poses an issue when trying to obtain the ``final'' of these three profiles because logistically, they are produced in sequence. The result is that the ``final'' cloud profile uses a pressure-temperature profile that is one iteration behind the ``final'' converged profile published in online repositories. In some cases, these two are negligibly different. However, in other cases the pressure-temperature profiles are different enough to inhibit perfect reproducibility of the cloud profiles from the Diamondback Zenodo repository \citep{morley_2024_12735103}. We demonstrate this subtlety in this benchmark test. 

We first take the pressure-temperature profiles, net fluxes, and $K_{\rm{zz}}$ profiles from the original climate model log files (EGP, in this case) generated for the Diamondback grid, which are available upon request  (priv. comm. C. Morley). As described in \S \ref{sec:kzz}, when the $K_{\rm{zz}}$ profile is computed within EGP it uses the convective heat flux and H$_2$ adiabatic profile. When running \texttt{Virga} standalone, we can also compute the $K_{\rm{zz}}$ by directly inputting this convective heat flux. In the absence of the convective heat flux, we can approximate it by assuming it scales as $\sigma T^4$. The comparison of these three is shown in Figure \ref{fig:diamondback1}A: 1) raw $K_{\rm{zz}}$ from EGP log files, 2) $K_{\rm{zz}}$ computed from a user input convective heat flux in \texttt{Virga}, and 3) $K_{\rm{zz}}$ computed with \texttt{Virga} by using temperature to approximate convective heat flux. If given the realistic convective heat flux, \texttt{Virga} can reproduce the  $K_{\rm{zz}}$ profile from EGP. However, using an approximation via the temperature leads to a $K_{\rm{zz}}$ that differs from the EGP values by up to two orders of magnitude. 

Next, we compute layer optical depth, asymmetry, and single scattering albedos with \texttt{Virga} using the $K_{\rm{zz}}$ profiles shown in Figure~\ref{fig:diamondback1}A and the Diamondback modeling setup. Figure~\ref{fig:diamondback1}C, D, and E shows the comparison between those three $K_{\rm{zz}}$ profiles and the Diamondback cloud profiles using the exact pressure-temperature profile from the EGP log files (not the ``final'' profile from Zenodo). These figures demonstrate our ability to reproduce the Diamondback optical properties. They also demonstrate how the crude estimate for $K_{\rm{zz}}$ impacts the optical properties. Users should not assume they can estimate $K_{\rm{zz}}$ with an $\sigma$T$^4$ estimate of the convective heat flux. Finally, Figure~\ref{fig:diamondback1}B demonstrates how the reproducibility of the cloud optical properties translates to reproducibility of the spectra using \texttt{PICASO} v3.3 \citep{Batalha2019ApJ...878...70B,Mukherjee2023ApJ...942...71M} with Zenodo v3 opacities \citep{natasha_batalha_2025_14861730}.  

This near exact matching example shown in Figure~\ref{fig:diamondback1} relies on unpublished pressure-temperature profiles. Figure~\ref{fig:diamondback2}, in contrast, shows the cloud optical depth profile computed with the published Zenodo pressure-temperature profiles. In the T=1900~K case, the optical depth profiles look qualitatively similar with minor differences. In the T=900~K case, the final converged pressure-temperature profile produces a cloud base that is lower in pressure by $\sim0.5$~dex bar. Note that ultimately though, the high optical depth regions agree well. Therefore, users are still encouraged to take the Diamondback Zenodo pressure-temperature profiles and experiment with \texttt{Virga} to test out varying parameters not explored in the Diamondback grid. These include variables such as particle distribution width, additional $K_{\rm{zz}}$ profiles, or using a different  $f_\text{sed}$ prescription.

\section{Discussion \& Conclusion}\label{sec:discon}
We have presented the official documentation for the open-source package \texttt{Virga}, which is based on the widely adopted methodology of \citet{ackerman2001cloud}. We provided an overview of the code workflow itself, describing how the code splits the atmosphere into layers, then sub-layers to converge on solutions for particle radii, and condensate mixing ratios. We also described how those calculations are used to compute final atmospheric scattering properties such as layer optical depth, single scattering albedo and asymmetry. 

Furthermore, we provided in depth discussion of updates made since the original publication of \citet{ackerman2001cloud} and follow-up of \citet{morley2012neglected}. This includes updates to the creation of the back-end optical constants data, calculations of the Mie properties, the total number of available condensate species, the saturation vapor pressure curves, and lastly the methodology for the fall speed calculations. For the latter, we provide an alternative method for the calculation of fall speeds based on the work of \citet{khan1987resistance}, which allows us to use a single expression for the drag coefficient and mitigates discontinuities between terminal velocity regimes. For this version of the code we leave this as extra functionality, but will considering making it the default formalism in a subsequent version of \texttt{Virga}.

Lastly, we benchmark \texttt{Virga} against a published use-case in both exoplanets and brown dwarfs. First, we reproduce the SiO$_2$ cloud detection results of \citet{Grant+2023} and show how the spectral model may have differed if we had blindly condensed all gases that are available to condense via \texttt{Virga}'s recommend gas function. We urge users of \texttt{Virga} to be mindful of what gases they include in their modeling as it should depend on observational context and the specific scientific use case. Next, we show how \texttt{Virga-v1} can be used to reproduce the cloud models of the Sonora Diamondback grid, which is pertinent to brown dwarfs with effective temperatures from 900 to 2400 K. 

Development of \texttt{Virga} is ongoing. Version 2, which focuses on modeling of aggregate particles (Moran and Lodge et al. 2025, submitted) is already staged to be released. Future versions are in planning stages. The authors encourage collaborative development of this open-source code and encourage community members to open issues, recommend feature improvements, and submit code modifications.

\begin{acknowledgments}
N.E.B acknowledges support from NASA’S Interdisciplinary Consortia for Astrobiology Research (NNH19ZDA001N-ICAR) under award number 19-ICAR19\_2-0041. C.R.’s research was supported by an appointment to the NASA Postdoctoral Program at the NASA Ames Research Center, administered by Universities Space Research Association under contract with NASA.

\end{acknowledgments}

\begin{contribution}
NE Batalha wrote the original Virga package, implemented critical fixes, and manages the code package. S Mukherjee wrote the internal Mie code that was used to benchmark the use of PyMieScatt and now MiePython. C Rooney added in new functionality such as variable fsed, the direct solver, and helped with benchmarking. C Visscher did all the calculations for the saturation vapor pressure curves and gas mean mass mixing ratios. SE Moran added the flex cloud, and user defined particle distribution functionality. MG Lodge implemented corrected Mie grid radius binning. M Marley provided the guidance needed to interpret the original fortran code, which Fortney helped to test. AR Sengupta implemented the comparison to the Diamondback grid. Mang, Mayorga, Morley, Kiefer all contributed to critical bug fixes and testing. All authors contributed to writing the manuscript.
\end{contribution}

\software{numba \citep{numba}, pandas \citep{mckinney2010data}, bokeh \citep{bokeh}, NumPy \citep{walt2011numpy}, IPython \citep{perez2007ipython}, Jupyter \citep{kluyver2016jupyter}, PySynphot \citep{pysynphot2013}, PICASO \citep{picaso1}, Virga \citep{virga1} }

\end{document}

%% file: diagram.tex
\begin{tikzpicture}[node distance=2cm]

\node (start) [startstop] {Start};
\node (mie) [io, below of=start] {\texttt{get\_mie}: 
Loads Mie database for each requested condensate on a grid of Q$_\mathrm{ext}$, Q$_\mathrm{scat}$, and $g$ for a user-defined radius and wavelength grid. 
};
\node (eddysed) [process2, below of=mie, yshift=-2cm] {\texttt{eddysed}: 
loops through each gas condensate and individual pressure layers. For each of these it calls a function called \texttt{layer} which returns the condensate mixing ratio,  $q_c$, which is needed as input for each recursive layer. It also stores the final computed particle radii and number density from  \texttt{layer} for each pressure layer and condensate.
};
\node (layer) [process2, right of=eddysed, xshift=4cm] {\texttt{layer}:
divide the original pressure layer grid into sub-layers in order to converge on a cloud layer optical depth. This is necessary because the optical depth has an inherent dependence on the vertical grid resolution. Once the final layer optical depth is converged, the condensate mixing ratio (Eqn.\ref{eqn:qc}), column droplet number concentration (Eqn.\ref{eqn:ndz}), and geometric radii (Eqn.\ref{eqn:rg}).
};
\node (calcqc) [process2, below of=layer, yshift=-2cm] {\texttt{calc\_qc}: 
computes condensate mixing ratio (Eqn.\ref{eqn:qt}), the column droplet number concentration (Eqn.\ref{eqn:ndzi}), and effective area-weighted condensate radius  (Eqn.\ref{eqn:reffi}), for a sub layer by assuming geometric scatterers. 
};
\node (calcopt) [process2, below of=eddysed, yshift=-2cm] {\texttt{calc\_opt}:
integrates pre-loaded Mie properties given lognormal-distribution-derived number density for the loaded radius grid and returns extinction optical depth, asymmetry, and single scattering albedo.
};
\node (out1) [io, below of=calcopt, yshift=-1cm] {Output: 
condensate mixing ratio, optical depth per layer, single scattering albedo, phase asymmetry, mean particle radius, droplet effective radius
};
\node (stop) [startstop, below of=out1] {Stop};

\draw [arrow] (start) -- (mie); 
\draw [arrow] (mie) -- (eddysed); 
\draw [arrow] (eddysed) -- (layer);
\draw [arrow] (layer) -- (calcqc);
\draw [arrow] (eddysed) -- (calcopt);
\draw [arrow] (calcopt) -- (out1);
\draw [arrow] (out1) -- (stop);
\end{tikzpicture}